\documentclass[aps,prd,twocolumn,showpacs,nofootinbib]{revtex4-2}

\usepackage{graphicx}
\usepackage{amsmath}
\usepackage{amssymb}
\usepackage{bm}
\usepackage{bbm}
\usepackage{color}
\usepackage{slashed}
\usepackage{soul}
\usepackage[colorlinks=true,linkcolor=blue,citecolor=blue, urlcolor=black]{hyperref}

\usepackage[normalem]{ulem}

\begin{document}

\title{QCD corrections of $e^+e^- \to J/\psi+c+\bar{c}$ using the principle of maximum conformality}

\author{Xu-Dong Huang$^1$}
\email{huangxd@cqnu.edu.cn}

\author{Xing-Gang Wu$^2$}
\email{wuxg@cqu.edu.cn}

\author{Xu-Chang Zheng$^2$}
\email{zhengxc@cqu.edu.cn}

\author{Bin Gong$^{3, 4}$}
\email{twain@ihep.ac.cn}

\author{Jian-Xiong Wang$^{3, 4}$}
\email{jxwang@ihep.ac.cn}

\affiliation{$^1$ College of Physics and Electronic Engineering, Chongqing Normal University, Chongqing 401331, P.R. China}
\affiliation{$^2$ Department of Physics, Chongqing Key Laboratory for Strongly Coupled Physics, Chongqing University, Chongqing 401331, P.R. China}
\affiliation{$^3$ Institute of High Energy Physics, Chinese Academy of Sciences, 19B Yuquan Road, Shijingshan District, Beijing, 100049, P.R. China}
\affiliation{$^4$ University of Chinese Academy of Sciences, Chinese Academy of Sciences, 19A Yuquan Road, Shijingshan District, Beijing, 100049, P.R. China}

\date{\today}

\begin{abstract}
In this paper, we compute the total and differential cross sections for $e^+e^- \to J/\psi+c+\bar{c}$ at the $B$ factories up to next-to-leading order (NLO) corrections within the framework of nonrelativistic QCD factorization theory. We then obtain improved pQCD series of those cross sections by using the Principle of Maximum Conformality (PMC). We show that the PMC can be applied for any pQCD calculable observable at the total and differential levels via a self-consistent way in perturbation theory. We observe that a more precise prompt total cross section at the NLO level can be achieved after applying the PMC, e.g. $\sigma|_{\rm prompt}^{\rm PMC}= 0.565^{+0.144}_{-0.125}~\text{pb}$. Here the uncertainty is the squared average of those from the $\alpha_s$ fixed-point uncertainty $\Delta\alpha_s(M_Z)$, the uncertainty of charm quark mass $\Delta m_c$, and an estimated contribution of the uncalculated NNLO-terms as predicted by the Pad\'{e} approximation approach. The differential cross sections $d\sigma/dP_{J/\psi}$, $d\sigma/d|\cos \theta|$, and $d\sigma/dz$ for $e^+e^- \to J/\psi+c+\bar{c}$ are further examined. Those results show that by further considering the feed-down contributions, the PMC predictions show better agreement with the Belle measurements.
\end{abstract}

\maketitle

\flushbottom

\section{Introduction}
\label{secIntro}

Heavy quarkonium production serves as an ideal laboratory for investigating the interplay between the perturbative and nonperturbative effects in Quantum Chromodynamics (QCD) theory. Since the discovery of $J/\psi$ in 1974, it has been a focal point of both theoretical and experimental researches. To characterize quarkonium production, various models have been introduced, including the colour-evaporation model (CEM)~\cite{Fritzsch:1977ay, Halzen:1977rs}, the color-singlet model (CSM)~\cite{Chang:1979nn, Berger:1980ni, Matsui:1986dk}, the nonrelativistic QCD (NRQCD) effective theory~\cite{Bodwin:1994jh}, and etc.. Among these, the NRQCD effective theory offers a systematic approach for distinguishing between short-distance and long-distance effects in quarkonium production. Notable successes have been achieved for interpreting the experimental data on quarkonium production, particularly in the context of unpolarized cross sections for $J/\psi$ hadroproduction, c.f. the reviews~\cite{Brambilla:2010cs, Andronic:2015wma, Lansberg:2019adr, Chen:2021tmf} and references therein. However, there are still challenges to NRQCD. For example, the hadroproduction cross section of $\eta_c$ as observed in the LHCb experiments~\cite{LHCb:2014oii} can be effectively explained by the color-singlet contribution alone, suggesting that the color-octet contribution should be very small~\cite{Butenschoen:2014dra}. This observation appears to conflict with the heavy-quark spin symmetry (HQSS) relation that exists between the long-distance matrix elements (LDMEs) of $\eta_c$ and $J/\psi$~\cite{Han:2014jya, Zhang:2014ybe, Belle:2009bxr, Li:2014fya}. Further exploration of processes involving charmonium is essential to test the NRQCD factorization.

A high luminosity $e^+ e^-$ collider has three general features such as cleanliness, democracy, and holism~\cite{Murayama:1996ec}, which are helpful and have some advantages in performing more precise measurements. The inclusive $J/\psi$ production cross section via the $e^+e^-$ annihilation has been measured by the BaBar collaboration as $2.54\pm0.21\pm0.21\mathrm{~pb}$~\cite{BaBar:2001lfi, BaBar:2005nic} and by the Belle collaboration as $1.45\pm0.10\pm0.13\mathrm{~pb}$~\cite{Belle:2001lqi, Belle:2002tfa}. Many theoretical investigations have been conducted on this production at leading-order (LO) within NRQCD, yielding results for inclusive $J/\psi$ production within the range of $0.6\sim1.7\mathrm{~pb}$ depending on parameter selections~\cite{Keung:1980ev, Driesen:1993us, Yuan:1996ep, Cho:1996cg, Schuler:1998az, Baek:1998yf, Liu:2002wq, Hagiwara:2007bq, Braaten:1995ez, Wang:2003fw}. A subsequent analysis by the Belle collaboration reports~\cite{Belle:2002tfa}
\begin{equation}
\sigma[e^+e^-\rightarrow J/\psi + c+\bar{c}]=0.87^{+0.21}_{-0.19}\pm 0.17\mathrm{~pb},
\end{equation}
and suggests $\sigma[e^+e^-\rightarrow J/\psi+X({\mathrm{non}~c\bar{c}})]\sim0.6\mathrm{~pb}$. Reference~\cite{Ioffe:2003gj} proposes that distinct $J/\psi$ production mechanisms can be discerned through measurements of $J/\psi$ polarization. Furthermore, a study on the $J/\psi$ polarization in $B\rightarrow J/\psi+ X$ employing the Belle detector is presented in Ref.~\cite{Ichizawa:2000yh}.

Regarding the $c\bar{c}$ component, the experimental data issued by the Belle collaboration~\cite{Belle:2002tfa}, $0.87^{+0.21}_{-0.19}\pm 0.17\mathrm{~pb}$, is approximately 5 times greater than the LO NRQCD prediction~\cite{Liu:2002wq}. This substantial deviation has been partially reconciled by considering both the next-to-leading-order (NLO) corrections and the feed-down contributions originating from the higher excited states~\cite{Zhang:2006ay, Gong:2009ng}. Additionally, it has been highlighted in Ref.~\cite{Nayak:2007mb} that the color transfer in associated heavy-quarkonium production may offer significant contributions to this process. The updated Belle experimental measurement~\cite{Belle:2009bxr} provides
\begin{equation}
\sigma[e^+e^-\rightarrow J/\psi +c+\bar{c}]=0.74\pm0.08^{+0.09}_{-0.08}\mathrm{~pb}, \label{belleII}
\end{equation}
which is closer to theoretical predictions. It has been shown that the NLO corrections are comparable with the LO corrections and are important to explain the data. Considering the significance of those corrections, it is imperative to undertake the computation of higher-order QCD corrections to yield more refined theoretical predictions. Nevertheless, the calculation of NNLO and higher-order QCD corrections to this process are quite difficult and are hard to be done in the near future. Thus, efforts directed towards offering a more accurate theoretical prediction at the NLO level are of significant importance.

A physical observable should be independent of the choices of renormalization scale and renormalization scheme. This is the central property of renormalization group invariance (RGI)~\cite{Stueckelberg:1953dz, GellMann:1954fq, Bogoliubov, Callan:1970yg, Symanzik:1970rt, Peterman:1978tb}. However there is renormalization scale ambiguity inherent in any initial fixed-order perturbative QCD (pQCD) series due to the mismatching of $\alpha_s$ and the coefficients at each perturbative order, resulting in a substantial theoretical uncertainty in pQCD predictions; e.g. it has been found that the initial NLO pQCD cross-sections for $e^+e^- \to J/\psi+c+\bar{c}$ at the $B$ factories are highly scale dependent. Therefore, it is crucial to eliminate the renormalization scale ambiguity to achieve a more accurate fixed-order prediction.

Many methods have been suggested to achieve the goal, among them the principle of maximum conformality (PMC)~\cite{Brodsky:2011ta, Brodsky:2011ig, Brodsky:2012rj, Mojaza:2012mf, Brodsky:2013vpa} has been suggested and developed since 2011. It suggests a systematic approach to achieve a scheme-and-scale invariant conformal series from the initial pQCD series. The PMC provides the underlying principle for the BLM method~\cite{Brodsky:1982gc} and extends its procedures from the NLO level to all orders. In the limit of small number of colors $N_c\to 0$~\cite{Brodsky:1997jk}, the PMC reduces to the Gell-Mann Low method~\cite{Gell-Mann:1954yli} for QED.

The PMC determines the correct magnitude of $\alpha_s$ by recursively using the renormalization group equation (RGE). It should be pointed out that two steps need to be done to achieve a correct PMC prediction. The first step is to get all the $\{\beta_i\}$-terms of the process, and the second step is to deal with those $\{\beta_i\}$-terms properly. That is, only the RGE-involved $\{\beta_i\}$-terms should be used to fix the correct magnitude of $\alpha_s$, and other $\{\beta_i\}$-terms which are connection to the quark-mass anomalous dimension $\gamma_m$-functions should be used to fix the running quark mass~\cite{Huang:2022rij, Ma:2024xeq}, and etc.. After applying the PMC, the resulting improved perturbative series becomes scale-invariant that is independent of any initial choice of renormalization scale~\cite{Wu:2018cmb, Huang:2021hzr}. Due to the elimination of all the scheme-dependent non-conformal $\{\beta_i\}$-terms, the resulting perturbative QCD series transforms into a scheme-independent conformal series~\cite{Wu:2013ei, Wu:2014iba, Wu:2019mky, DiGiustino:2023jiq}.

In this paper, by applying the PMC, we will provide a comprehensive and more accurate analysis of the NLO QCD correction to the production channel $e^+e^-\to J/\psi +c+\bar{c}$ at the $B$ factories with $e^+ e^-$ collision energy $\sqrt{s}=10.6$ GeV.

The remaining parts of the paper are organized as follows. In Sec.II, we give the LO cross section for the process. The calculation of NLO QCD corrections and the PMC procedures are described in Sec.III. Numerical results and discussions are presented in Sec.IV. Section V is reserved for a summary.

\section{LO cross section} \label{secLO}

At the $B$ factories, the $e^+ e^-$ will first annihilate into a virtual photon and then forms the wanted final state. Apart from the fact that $c$ and $\bar{c}$ quarks combine into $J/\psi$ non-perturbatively, the rest part is always ``hard" and can be calculated perturbatively. Within the framework of NRQCD factorization, the differential cross section for $e^+(k_1)+e^-(k_2) \to \gamma^* \to J/\psi(p_1)+c(p_2)+\bar{c}(p_3)$ can be expressed as
\begin{eqnarray}
d\sigma_{e^+e^- \to J/\psi+c+\bar{c}} =\sum_n d\hat{\sigma}_{e^+e^- \to (c\bar{c})[n]+c+\bar{c}}\langle {\cal O}^{J/\psi}(n)\rangle,\label{nrqcdfact}
\end{eqnarray}
where $d\hat{\sigma}$ represents the perturbatively calculable short-distance coefficients (SDCs) and $\langle {\cal O}^{J/\psi}(n)\rangle$, in which $n$ stands for the quantum numbers of the intermediate on-shell $(c\bar{c})$-pair, are long-distance matrix elements (LDMEs) which are non-perturbative but can be determined by fitting experimental data or estimated by using proper potential models. The summation encompasses all the possible intermediate color-singlet and color-octet states of the $(c\bar{c})$-pair, e.g. ${^{2S+1}L}_{J}^{[1,8]}$. In the lowest-order nonrelativistic approximation, only the color-singlet contribution with $n={^{3}S}_1^{[1]}$ needs to be considered.

\begin{figure}[htbp]
\center
\includegraphics[width=0.45\textwidth]{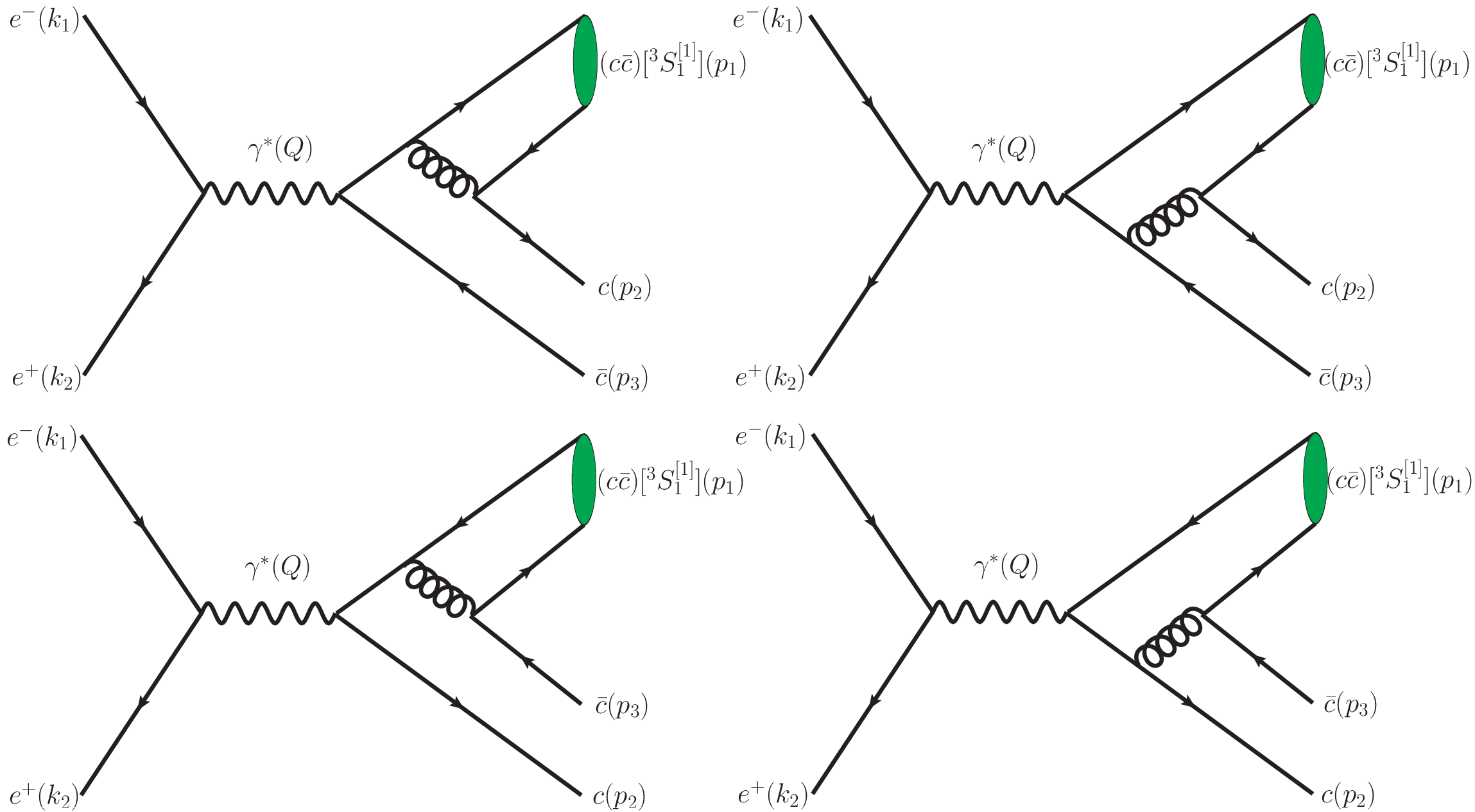}
\caption{The LO Feynman diagrams for $e^+e^- \to (c\bar{c})[n]+c+\bar{c}$ with $n={^{3}S}_1^{[1]}$.
 } \label{feylo}
\end{figure}

At the LO level, there are four Feynman diagrams for the process $e^+e^- \to (c\bar{c})[n]+c+\bar{c}$, as illustrated in Fig.~\ref{feylo}. Corresponding to these four Feynman diagrams, the LO amplitude for this process can be expressed as $M_{\rm LO}=M_1+M_2+M_3+M_4$, where
\begin{eqnarray}
iM_1&=&-\frac{-i}{(p_1/2+p_2)^2+i\epsilon}\frac{-i}{(p_1+p_2+p_3)^2+i\epsilon} \nonumber \\
&& \cdot \bar{v}_{m_e}(k_2)(ig \gamma^{\mu})u_{m_e}(k_1)\bar{u}_{m_c}(p_2)(-ig_s \gamma^{\nu}T^a) \nonumber \\
&& \cdot \Pi_1 \Lambda_1 (-ig_s \gamma_{\nu}T^a) \frac{i}{\slashed{p}_1+\slashed{p}_2-m_c+i \epsilon} \nonumber \\
&& \cdot \left(-\frac{2}{3}ig \gamma_{\mu}\right) v_{m_c}(p_3),
\end{eqnarray}
\begin{eqnarray}
iM_2&=&-\frac{-i}{(p_1/2+p_2)^2+i\epsilon}\frac{-i}{(p_1+p_2+p_3)^2+i\epsilon} \nonumber \\
&& \cdot \bar{v}_{m_e}(k_2)(ig \gamma^{\mu})u_{m_e}(k_1)\bar{u}_{m_c}(p_2)(-ig_s \gamma^{\nu}T^a) \nonumber \\
&& \cdot \Pi_1 \Lambda_1 \left(-\frac{2}{3}ig \gamma_{\mu}\right) \frac{i}{-\slashed{p}_1/2-\slashed{p}_2-\slashed{p}_3-m_c+i \epsilon} \nonumber \\
&& \cdot (-ig_s \gamma_{\nu}T^a)v_{m_c}(p_3),
\end{eqnarray}
\begin{eqnarray}
iM_3&=&-\frac{-i}{(p_1/2+p_3)^2+i\epsilon}\frac{-i}{(p_1+p_2+p_3)^2+i\epsilon} \nonumber \\
&& \cdot \bar{v}_{m_e}(k_2)(ig \gamma^{\mu})u_{m_e}(k_1)\bar{u}_{m_c}(p_2)\left(-\frac{2}{3}ig \gamma_{\mu}\right) \nonumber \\
&& \cdot \frac{i}{-\slashed{p}_1-\slashed{p}_3-m_c+i \epsilon}(-ig_s \gamma^{\nu}T^a)\Pi_1 \Lambda_1   \nonumber \\
&& \cdot (-ig_s \gamma_{\nu}T^a)v_{m_c}(p_3),
\end{eqnarray}
\begin{eqnarray}
iM_4&=&-\frac{-i}{(p_1/2+p_3)^2+i\epsilon}\frac{-i}{(p_1+p_2+p_3)^2+i\epsilon} \nonumber \\
&& \cdot \bar{v}_{m_e}(k_2)(ig \gamma^{\mu})u_{m_e}(k_1)\bar{u}_{m_c}(p_2)(-ig_s \gamma^{\nu}T^a) \nonumber \\
&& \cdot \frac{i}{\slashed{p}_1/2+\slashed{p}_2+\slashed{p}_3-m_c+i \epsilon} \left(-\frac{2}{3}ig \gamma_{\mu}\right) \Pi_1 \Lambda_1 \nonumber \\
&& \cdot (-ig_s \gamma_{\nu}T^a)v_{m_c}(p_3),
\end{eqnarray}
where, $\Pi_1$ is the projector for the spin-triplet $S$-wave state:
\begin{eqnarray}
\Pi_1=\frac{1}{(2\,m_c)^{3/2}}(\slashed{p}_1/2-m_c)\slashed{\epsilon}^{*}(p_1)(\slashed{p}_1/2+m_c),
\end{eqnarray}
and $\Lambda_1$ is the color projector for the color-singlet state:
\begin{eqnarray}
\Lambda_1=\frac{\textbf{1}}{\sqrt{3}},
\end{eqnarray}
where $\textbf{1}$ represents the unit matrix of the $SU(3)_c$ group.

Using the above amplitudes, the LO differential cross section for $e^+e^- \to (c\bar{c})[n]+c+\bar{c}$ can be computed by
\begin{eqnarray}
d\sigma_{\rm LO}=\frac{1}{4}\frac{1}{2s}\sum \vert M_{\rm LO} \vert^2 d\Phi_3,
\end{eqnarray}
where $s=(k_1+k_2)^2$, $1/4$ comes from the spin average of the incident electron and positron, $1/(2s)$ is the flux factor, and $\sum$ denotes the summation over the spin and color states of the final-state particles. $d\Phi_3$ represents the differential phase space for the three-body final state, and
\begin{eqnarray}
d\Phi_3=(2\pi)^D\delta^D\left(\sum_{j=1}^2 k_j-\sum_{f=1}^3 p_f\right)\prod_{f=1}^3 \frac{d^{D-1} \textbf{p}_f}{(2\pi)^{D-1} 2E_f},\label{dphi3}
\end{eqnarray}
where $D$ represents the space-time dimension. With these formulas, the LO cross section for $e^+e^- \to (c\bar{c})[n]+c+\bar{c}$ can be calculated directly.

\section{NLO corrections} \label{secNLO}

At the NLO level, the virtual and real corrections need to be calculated. The virtual correction exhibits ultraviolet (UV) and infrared (IR) divergences, while the real correction presents an IR divergence. In this paper, the conventional dimensional regularization with $D=4-2\epsilon$ is employed to regularize both UV and IR divergences. Then the UV and IR divergences appear as pole terms in $1/\epsilon$. In the subsequent subsections, we elaborate on the key steps involved in calculating the virtual and real corrections.

\subsection{Virtual NLO correction}

\begin{figure}[htbp]
\center
\includegraphics[width=0.45\textwidth]{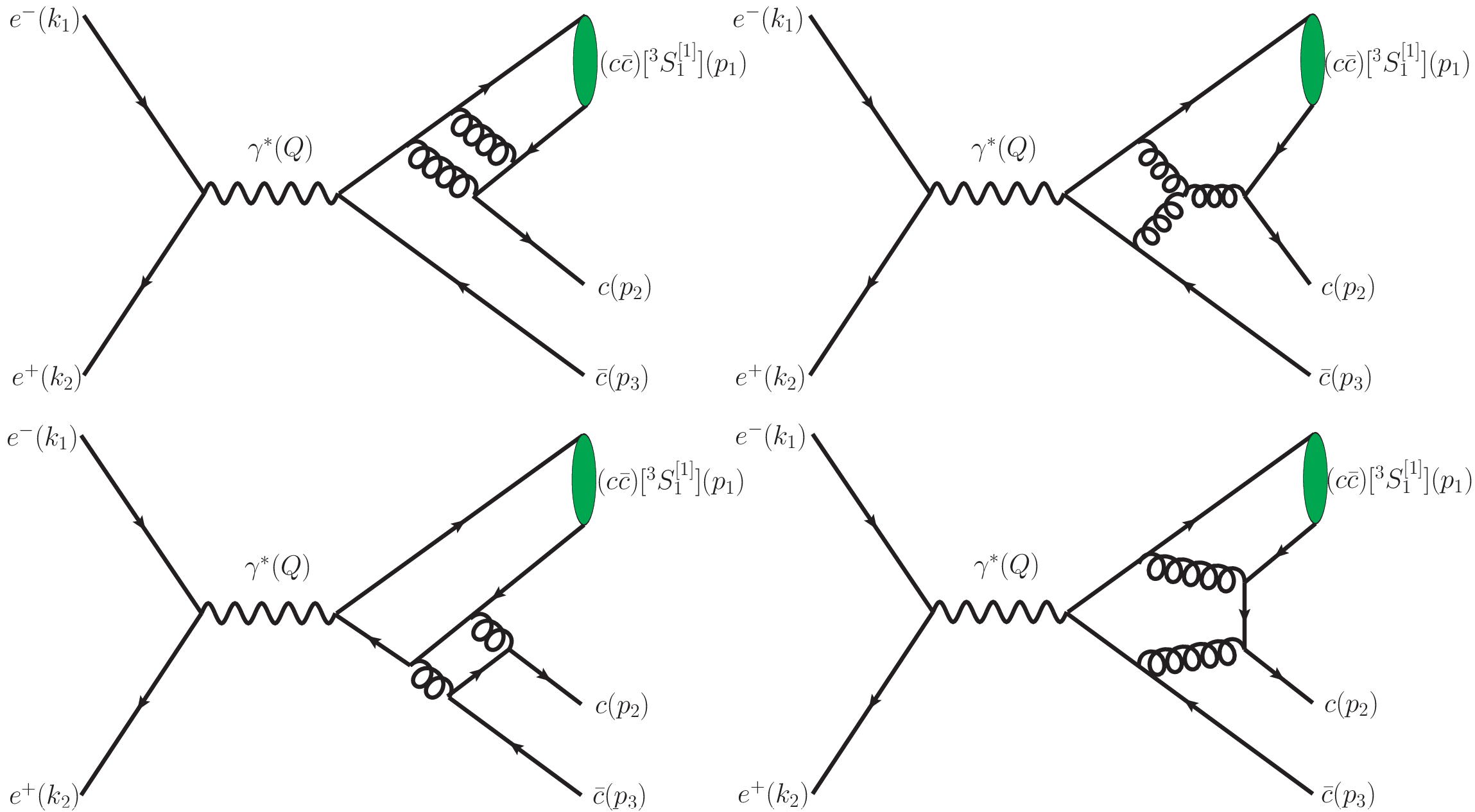}
\caption{Four typical one-loop Feynman diagrams for $e^+e^- \to (c\bar{c})[n]+c+\bar{c}$ with $n={^{3}S}_1^{[1]}$.} \label{feyvir}
\end{figure}

The virtual correction at the NLO level arises from the interference between the one-loop Feynman diagrams and the tree Feynman diagrams. There are totally 72 Feynman diagrams for the virtual correction. Four representative one-loop Feynman diagrams are depicted in Fig.~\ref{feyvir}.

The virtual correction to the cross section of the process $e^+e^- \to (c\bar{c})[n]+c+\bar{c}$ can be calculated through
\begin{equation}
d\sigma_{\rm Virtual}=\frac{1}{4}\frac{1}{2s}\sum 2 {\rm Re}\left(M^*_{\rm LO} M_{\rm Virtual} \right)d\Phi_3,
\end{equation}
where $M_{\rm Virtual}$ represents the amplitude for the virtual correction.

Feynman diagrams with a virtual gluon line connected to the quark pair in a meson exhibit Coulomb singularity, which manifests as power divergences in the IR limit of relative momentum. Such singularity can be addressed through $(c\bar{c})[n]$ wave function renormalization~\cite{Kramer:1995nb, Gong:2007db}. As a result, no Coulomb divergence appears in the resultant pQCD series. In dimensional regularization, there is a simpler way to extract the NRQCD short-distance coefficients using the method of regions~\cite{Beneke:1997zp}. In this method, one can calculate the hard region contributions directly by expanding the relative momentum of the $(c\bar{c})[n]$-pair before carrying out the loop integration. More explicitly, for the present case, one can directly set the relative momentum $q$ to zero before performing loop integration. The Coulomb divergence vanishes during the calculation with dimensional regularization. Moreover, the IR divergences originating from the virtual correction will be cancelled by the IR divergences stemming from the real correction. The UV divergences need to be removed via the process of renormalization. The renormalization scheme is as follows: the on-shell (OS) scheme is employed for the renormalization of the heavy quark field, the heavy quark mass, and the gluon field; the modified minimal subtraction ($\overline{\rm MS}$) scheme is adopted for the renormalization of the strong coupling constant. Using renormalization schemes, the quantities $\delta Z_i \equiv Z_i-1$ can be derived, e.g.
\begin{eqnarray}
 \delta Z^{\rm OS}_{2}&=&-C_F \frac{\alpha_s}{4\pi}\left[\frac{1}{\epsilon_{\rm UV}}+ \frac{2}{\epsilon_{\rm IR}}-3~\gamma_E+3~ {\rm ln}\frac{4\pi \mu_R^2}{m_c^2}+4\right], \nonumber\\
\delta Z^{\rm OS}_{m}&=&-3~C_F \frac{\alpha_s}{4\pi}\left[\frac{1}{\epsilon_{\rm UV}}- \gamma_E+
 {\rm ln}\frac{4\pi \mu_R^2}{m_c^2}+\frac{4}{3}\right],\nonumber\\
 \delta Z^{\rm OS}_3&=&\frac{\alpha_s}{4\pi}\left[(\beta'_0-2C_A) \left(\frac{1}{\epsilon_{\rm UV}}-\frac{1}{\epsilon_{\rm IR}}\right) \right. \nonumber\\
 &&\left. - \frac{4}{3}T_F \left(\frac{1}{\epsilon_{\rm UV}}-\gamma_E + {\rm ln}\frac{4\pi \mu_R^2}{m_c^2}\right)\right], \nonumber\\
 \delta Z^{\overline{\rm MS}}_g&=&- \frac{\beta_0}{2}\frac{\alpha_s}{4\pi}\left[\frac{1}{\epsilon_{\rm UV}}- \gamma_E+ {\rm ln}~(4\pi) \right],
\end{eqnarray}
where $\mu_R$ is the renormalization scale, $\gamma_E$ is the Euler constant. $\beta_0=11C_A/3-4T_F n_f/3$ is the one-loop coefficient of the QCD $\beta$ function, in which $n_f$ is the number of active quark flavors. $\beta'_0=11C_A/3-4T_F n_{lf}/3$ and $n_{lf}=3$ is the number light-quark flavors. For $SU(3)_c$ group, $C_A=3$, $C_F=4/3$ and $T_F=1/2$.

\subsection{Real NLO correction}

\begin{figure}[htbp]
\center
\includegraphics[width=0.45\textwidth]{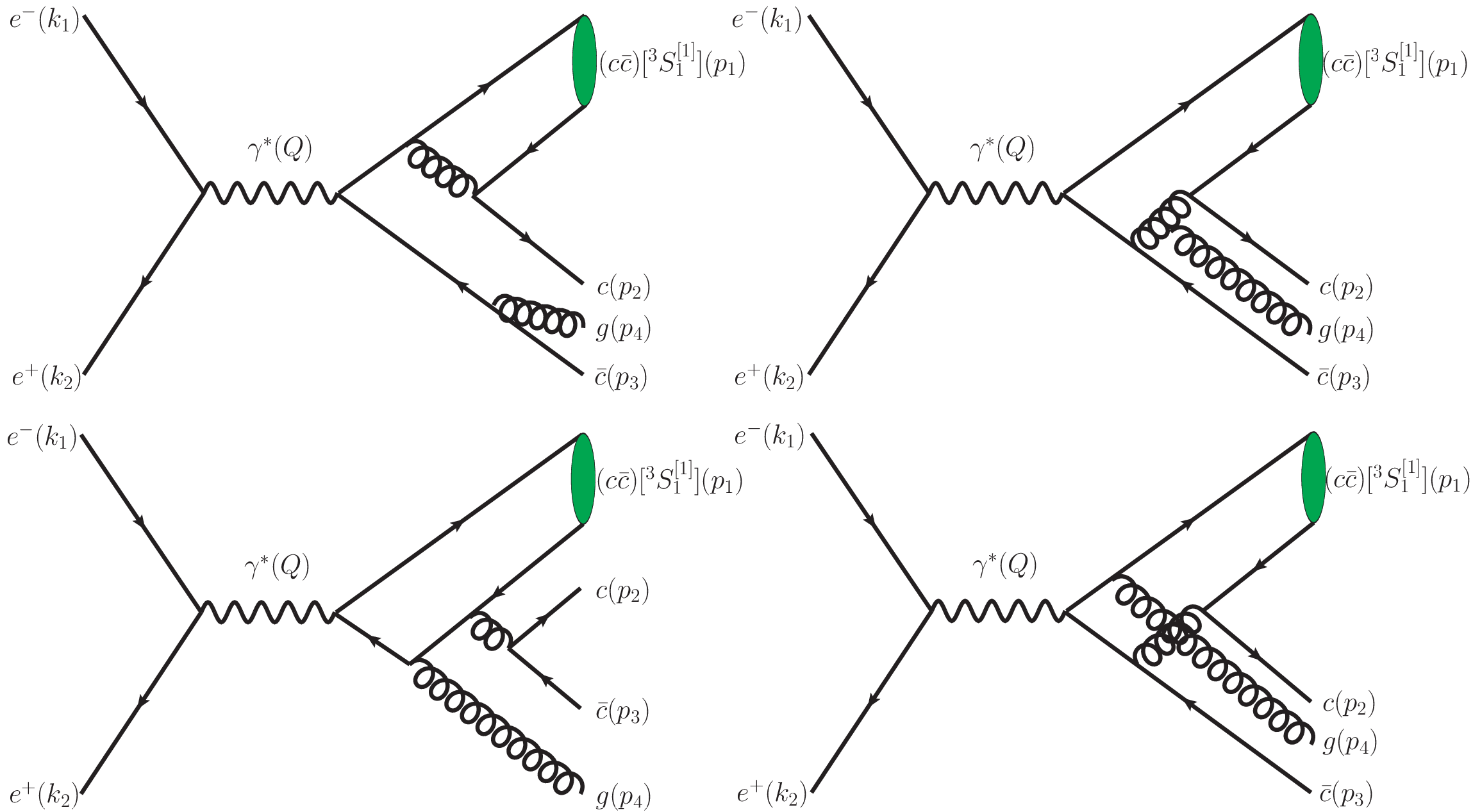}
\caption{Four typical real-correction Feynman diagrams for the decay process, $e^+e^- \to (c\bar{c})[n]+c+\bar{c}$ with $n={^{3}S}_1^{[1]}$
 } \label{feyreal}
\end{figure}

The real correction to the process $e^+e^- \to (c\bar{c})[n]+c+\bar{c}$ arises from the process $e^+(k_1)+e^-(k_2) \to (c\bar{c})[n](p_1) +c(p_2) +\bar{c}(p_3) +g(p_4)$. There are totally 30 Feynman diagrams for the real correction. Four typical Feynman diagrams illustrating this process are depicted in Fig.~\ref{feyreal}. With these Feynman diagrams, the amplitude ($M_{\rm Real}$) for the real correction can be directly formulated. Subsequently, the differential cross section for the real correction can be calculated through
\begin{eqnarray}
d\sigma_{\rm Real}=\frac{1}{4}\frac{1}{2s}\sum \vert M_{\rm Real} \vert^2 d\Phi_4,
\end{eqnarray}
where $d\Phi_4$ is the differential four-body phase space,
\begin{eqnarray}
d\Phi_4=(2\pi)^D\delta^D\left(\sum_{j=1}^2 k_j-\sum_{f=1}^4 p_f\right)\prod_{f=1}^4 \frac{d^{D-1} \textbf{p}_f}{(2\pi)^{D-1} 2E_f}.
\end{eqnarray}

The real correction exhibits IR divergences, originating from the phase-space integration in the region where the momentum of the final gluon is close to zero. To address these IR divergences in the real correction, we employ the two-cutoff phase-space slicing method~\cite{Harris:2001sx}. Due to the absence of collinear divergence in the current process, we introduce a single cutoff parameter denoted as $\delta_s$. Consequently, the phase space for the real correction is partitioned into two distinct regions: the soft region characterized by $E_4 \leq \sqrt{s} \delta_s/2$ and the hard region where $E_4 > \sqrt{s} \delta_s/2$. Here, the gluon energy $E_4$ is defined in the center-of-mass (CM) frame of the initial state particles $e^+e^-$. More explicitly, the real correction can be divided into two parts
\begin{eqnarray}
d\sigma_{\rm Real}=d\sigma_{\rm Real}^{\rm S}+d\sigma_{\rm Real}^{\rm H}.
\end{eqnarray}

By employing the soft approximation to both the amplitude and the phase space, we derive the contribution originating from the soft region, expressed as
\begin{eqnarray}
&&\hspace{-1em}d\sigma_{\rm Real}^{\rm S} \nonumber\\
&&\hspace{-1em}=d\sigma_{\rm LO}\left[\frac{C_F \alpha_s \Gamma(1+\epsilon)}{\pi}\left(\frac{4\pi\mu_R^2}{s}\right)^{\epsilon}\right] \left\{\left( \frac{1}{\epsilon}-{\rm ln}\,\delta_s^2 \right) \right. \nonumber\\
&&\hspace{-1em}~\times \left(1-\frac{\kappa\, p_2 \cdot p_3}{( \kappa^2-1)m_c^2}{\rm ln}\,\kappa^2\right)+\frac{1}{2\beta_2}{\rm ln}\left(\frac{1+\beta_2}{1-\beta_2}\right)\nonumber\\
&&\hspace{-1em}~+\frac{1}{2\beta_3}{\rm ln}\left(\frac{1+\beta_3}{1-\beta_3} \right) +\frac{2\,\kappa\, p_2 \cdot p_3}{( \kappa^2-1)m_c^2}\left[\frac{1}{4}{\rm ln}^2\left(\frac{1-\beta_2}{1+\beta_2} \right) \right. \nonumber \\
&&\hspace{-1em}~+{\rm Li}_2\left( 1-\frac{\,\kappa\, E_2(1+\beta_2)}{v}\right)  +{\rm Li}_2\left( 1-\frac{\,\kappa\, E_2(1-\beta_2)}{v}\right)\nonumber \\
&&\hspace{-1em}~-\frac{1}{4}{\rm ln}^2\left(\frac{1-\beta_3}{1+\beta_3} \right)-{\rm Li}_2\left( 1-\frac{E_3(1+\beta_3)}{v}\right)\nonumber \\
&&\hspace{-1em}~\left.\left. -{\rm Li}_2\left( 1-\frac{ E_3(1-\beta_3)}{v}\right)\right]\right\},
\end{eqnarray}
where $\beta_2=\sqrt{1-m_c^2/E_2^2}$, $\beta_3=\sqrt{1-m_c^2/E_3^2}$, $v= (\kappa^2-1)m_c^2/[2(\kappa\, E_2-E_3)]$, and $\kappa= [p_2\cdot p_3+\sqrt{(p_2 \cdot p_3)^2-m_c^4}]/m_c^2$. $E_2$ and $E_3$ are also defined in the CM frame of the initial-state particles $e^+e^-$.

Due to the constraint $E_4 > \sqrt{s}\delta_s/2$ for the hard region, the contribution stemming from this region is finite. Consequently, the contribution from the hard region can be computed numerically in four-dimensions. The real correction can be easily obtained by summing the contributions from the hard and soft regions. Both the contributions from the soft and hard regions depend on the cutoff parameter $\delta_s$ individually, but the sum of these two contributions should be independent of the choice of $\delta_s$. Verifying this $\delta_s$ independence is a crucial test of the accuracy of the calculation. We have confirmed the $\delta_s$ independence by observing that the results remain consistent across a range of $\delta_s$ values from $10^{-5}$ to $10^{-7}$. For definiteness, we will set the cutoff at $\delta_s=10^{-6}$ for the following numerical calculations.

The NLO corrections are obtained by summing the virtual and real corrections. Upon combining these corrections, the UV and IR divergences exactly cancelled, yielding finite results. The cross section $d\sigma_{e^+e^-\to J/\psi+c+\bar{c}}$ can be derived from $d\sigma_{e^+e^-\to (c\bar{c})[n]+c+\bar{c}}$ by multiplying it by a factor $\langle {\cal O}^{J/\psi}(n) \rangle/\langle {\cal O}^{(c\bar{c})[n]}(n) \rangle \approx \vert R_{J/\psi}(0)\vert^2/(4\pi)$, where $R_{J/\psi}(0)$ represents the $J/\psi$ radial wave function at the origin, which can be extracted from the leptonic decay widths $\Gamma(J/\psi \to e^+e^-)$.

In doing the calculations, we employ the package \texttt{FeynArts}~\cite{Hahn:2000kx} to generate Feynman diagrams and the related amplitudes, and the package \texttt{FeynCalc}~\cite{Mertig:1990an,Shtabovenko:2016sxi} to perform color and Dirac traces. The packages \texttt{\$Apart}~\cite{Feng:2012iq} and \texttt{FIRE}~\cite{Smirnov:2008iw} are employed for partial fraction and integration-by-parts (IBP) reduction. Following the IBP reduction, all one-loop integrals are reduced to master integrals, which are numerically evaluated using the package \texttt{LoopTools}~\cite{Hahn:1998yk}. The final phase-space integrations are computed with assistance from the package \texttt{Vegas}~\cite{Lepage:1977sw} and its alteration that can be found in the $B_c$ meson generator BCVEGPY~\cite{Chang:2003cq}.

\subsection{The NLO cross sections and the PMC}

Combing the above mentioned virtual and real contributions, the total cross section of $e^+(k_1)+e^-(k_2) \to \gamma^* \to J/\psi(p_1)+c(p_2)+\bar{c}(p_3)$ up to NLO QCD corrections can be expressed as
\begin{eqnarray}
\sigma_{\rm NLO}= c_{1}(\mu_R) a_s^2(\mu_R)+c_{2}(\mu_R) a_s^3(\mu_R), \label{nf}
\end{eqnarray}
where $a_s(\mu_R)=\alpha_s(\mu_R)/(4\pi)$. The coefficients $c_{1}$ and $c_{2}$ correspond to the LO-level and NLO-level QCD corrections, respectively. The LO-level coefficient $c_{1}$ has no $\mu_R$ dependence, and the NLO-level coefficient $c_{2}$ can be further decomposed into $n_f$-dependent and $n_f$-independent parts, i.e.,
\begin{eqnarray}
c_{2}(\mu_R)=c_{2,0}(\mu_R)+c_{2,1}(\mu_R) n_f, \label{c2}
\end{eqnarray}
where $n_f$ represents the number of active flavors, which is related to the QCD $\beta_0$-function through $\beta_0= 11 - 2/3 n_f$. The perturbative coefficients can be divided into conformal and non-conformal terms~\cite{Mojaza:2012mf, Brodsky:2013vpa}, and then the NLO total cross section (\ref{nf}) transforms to
\begin{eqnarray}
\sigma_{\rm NLO}= r_{1,0} a_s^2(\mu_R)+[r_{2,0}+2r_{2,1}\beta_0]a_s^3(\mu_R), \label{beta}
\end{eqnarray}
where the LO and NLO conformal coefficients $r_{1,0}=c_1$ and $r_{2,0}=c_{2,0}+33 c_{2,1}/2$, and the NLO non-conformal coefficient $r_{2,1}=-3c_{2,1}/4$.

The PMC provides a process-independent manner to fix conventional scheme-and-scale ambiguities. In practice, it is a kind of resummation, e.g. it resums all the known type of $\beta$-terms with the help of RGE and determines the correct magnitude of $\alpha_s$, whose precision is determined by the given order of the perturbative series. As for the present case, following the standard PMC scale-setting procedures, the NLO-level non-conformal $\beta_0$-terms will be resummed to achieve an effective strong coupling $a_s(Q_*)$, i.e.,
\begin{eqnarray}
a^2_s(Q_*) &\leftarrow& a^2_s(\mu_R)\left[1 + 2 \beta_0 \frac{r_{i+1,1}}{r_{i,0}} a_s(\mu_R) +\cdots \right],
\end{eqnarray}
which is derived by using the RGE. The scale $Q_*$ is usually called as the PMC scale and it can be determined at the leading-log (LL) accuracy
\begin{eqnarray}
Q_*=\mu_R {\rm exp} \left[-\frac{ r_{2,1}}{2 r_{1,0}}\right]. \label{Qstar}
\end{eqnarray}
The initial series (\ref{beta}) then transforms into the following series,
\begin{equation}
\sigma_{\rm NLO}^{\rm PMC}= r_{1,0} a_s^2(Q_*)+r_{2,0} a_s^3(Q_*). \label{PMC}
\end{equation}
Since the non-conformal terms have been removed, the improved series (\ref{PMC}) becomes scheme independent. Since the PMC scale $Q_*$ and the conformal coefficients $r_{(1,2), 0}$ are independent of the choice of $\mu_R$, such series (\ref{PMC}) is also scale independent. Thus, the present NLO total cross section (\ref{PMC}) provides another good application of PMC.

The above PMC procedures for the total cross section are also applicable for differential cross sections, such as the $J/\psi$ distributions for the kinematic parameters $P_{J/\psi}$, $|\cos\theta|$ and the energy fraction $z = 2(k_1+k_2) \cdot p_{1}/s$. Here $P_{J/\psi}$ and $\theta$ are $J/\psi$ three-momentum and $J/\psi$ production angle in the laboratory frame, respectively. By applying the PMC for the differential cross section, one may obtain a PMC scale for each variable such as $P_{J/\psi}$, $|\cos\theta|$ or $z$, which after integrating over the corresponding variable may result as the overall PMC scale in the sense of a mean value theorem. And by comparing the integration of those differential cross sections with the total cross section (\ref{PMC}), one may obtain a self-consistency check of the PMC scale-setting approach.

\section{Numerical results and discussions} \label{secNumer}

To do the numerical calculation, we adopt
\begin{eqnarray}
&& m_c=1.5\pm0.1\,{\rm GeV},\; \sqrt{s}=10.6\,{\rm GeV},\nonumber \\
&& \alpha=1/137,\vert R_{J/\psi}(0)\vert^2=0.918\,{\rm GeV}^{3},
\end{eqnarray}
where the electromagnetic coupling constant $\alpha$ is taken from Ref.~\cite{Gong:2009ng}. The input value for $\vert R_{J/\psi}(0)\vert^2$ is obtained by using $\vert R_{J/\psi}(0)\vert^2=\Gamma(J/\psi \to e^+e^-)9 m^2_{J/\psi}/[16\alpha^2(1-4C_F \alpha_s/\pi)]$, where $m_{J/\psi}=2m_c$ and $\Gamma(J/\psi \to e^+e^-)=5.53~{\rm KeV}$~\cite{ParticleDataGroup:2022pth}. For other values of $m_c$, the value of $\vert R_{J/\psi}(0)\vert^2$ should be multiplied by $(m_c/1.5~{\rm GeV})^2$. We use the package \texttt{RunDec3}~\cite{Herren:2017osy} to evaluate the running of the QCD coupling constant $\alpha_s(\mu_R)$ at two-loop accuracy. According to $\alpha_s(M_Z)=0.1179$~\cite{ParticleDataGroup:2022pth}, we obtain $\alpha_s(2m_c)=0.252$.

\subsection{Total cross section}

In this subsection, we give the numerical result for the total cross section of $e^+e^-\to J/\psi+c+\bar{c}$ up to NLO QCD corrections.

In order to have a glance at the size of the NLO QCD corrections, we first present the total cross section when the charm quark mass is taken as $m_c=1.5~{\rm GeV}$. The initial NLO series for the total cross section of $e^+e^-\to J/\psi+c+\bar{c}$ is
\begin{eqnarray}
\sigma_{\rm NLO}=c_1 a_s^2(\mu_R)+\bigg[c_2(2m_c) +2c_1\beta_0\ln\frac{\mu_R^2}{4m_c^2}\bigg]a_s^3(\mu_R), \label{conv}
\end{eqnarray}
where $c_1=391.589$ and $c_2(2m_c)=16231-580.686 n_f$. Starting from the scale-dependent initial series (\ref{conv}), one conventionally sets $\mu_R=2m_c$ to achieve its central value and varies $\mu_R\in [m_c, 4m_c]$ to estimate its scale uncertainty. For convenience, we call the results under this choice as conventional ones. After applying the PMC, following the above procedures~(\ref{nf}-\ref{Qstar}), we obtain the following scale invariant series:
\begin{eqnarray}
\sigma_{\rm NLO}^{\rm PMC}=391.589a_s^2(Q_*)+6649.681a_s^3(Q_*), \label{pmc}
\end{eqnarray}
where the LL-accuracy PMC scale, $Q_*=1.72~{\rm GeV}$.

\begin{figure}[htbp]
\center
\includegraphics[width=0.45\textwidth]{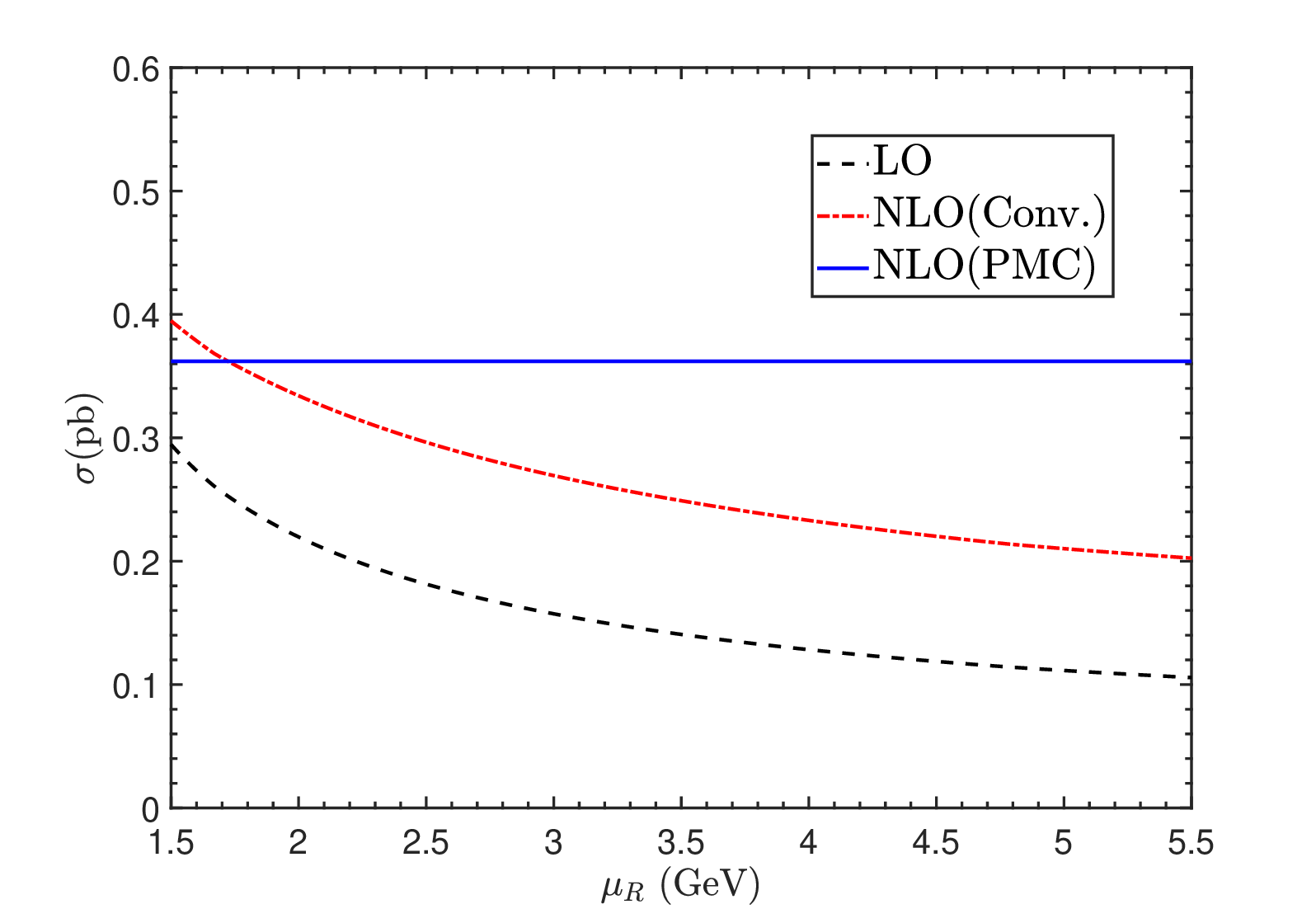}
\caption{The dependence of the LO and NLO total cross sections on the renormalization scale $\mu_R$. The conventional (Conv.) and PMC LO total cross sections are highly scale dependent and are the same. The conventional NLO total cross section is till scale dependent, but the PMC NLO prediction becomes scale independent. }
\label{urdependence}
\end{figure}

\begin{table}[htb]
\center
\begin{tabular}{c c c c}
\hline
  & ~~$\alpha_s^2$-terms~~ & ~~$\alpha_s^3$-terms~~ & ~~Total~~  \\
\hline
${\rm Conv.}$    & $0.157^{+0.137}_{-0.056}$ & $0.112^{-0.012}_{-0.017}$ & $0.269^{+0.125}_{-0.073}$    \\
${\rm PMC}$  & $0.253$ & $0.109$ & $0.362$  \\
\hline
\end{tabular}
\caption{Contributions of the $\alpha_s^2$-terms and the $\alpha_s^3$-terms to the NLO total cross sections (in unit: pb) of $e^+e^-\to J/\psi+c+\bar{c}$ under conventional (Conv.) and PMC scale-setting approaches, respectively. The PMC predictions are scale invariant. The conventional results are scale dependent, and their central values are for $\mu_R=2m_c$ and their errors are for $\mu_R\in [m_c, 4 m_c]$.}
\label{tb.sec}
\end{table}

We show how the LO and NLO total cross sections change with the renormalization scale $\mu_R$ in Fig.~\ref{urdependence}~\footnote{Here as a cross-check of our predictions for the initial NLO series, we have found that if taking the same values for the input parameters, we will obtain exactly the same conventional predictions of Refs.~\cite{Zhang:2006ay, Gong:2009ng}.}. The LO terms of conventional series are highly scale dependent. By including the NLO QCD corrections, the scale dependence of conventional series becomes smaller due to the cancellation of scale dependence among different orders, which is however still sizable, e.g. the net scale error is still $\sim 74\%$ for $\mu_R\in [m_c, 4 m_c]$. As for the PMC prediction, it becomes scale independent by using the $\{\beta_0\}$-terms to set the correct magnitude of $\alpha_s$. We also give separate contributions of the $\alpha_s^2$-terms and the $\alpha_s^3$-terms to the NLO total cross sections under conventional (Conv.) and PMC scale-setting approaches in Table \ref{tb.sec}. Table~\ref{tb.sec} shows that the $\alpha_s^3$-terms contribute significantly in both two cases, emphasizing the importance of the NLO-terms. The relative importance of $\alpha_s^2$-terms and $\alpha_s^3$-terms is $1: 71\%$ for $\mu_R=2 m_c$ under conventional scale-setting approach, which changes to $1: 43\%$ under PMC scale-setting approach. Thus the PMC series not only removes the scale dependence but also has a better pQCD convergence, showing the importance of using a proper scale-setting approach.

After eliminating the renormalization scale uncertainty via using the PMC approach, there are still several uncertainty sources, such as the contributions from the uncertainty of charm quark mass $\Delta m_c$, the $\alpha_s$ fixed-point uncertainty $\Delta\alpha_s(M_Z)$, the uncalculated higher-order (UHO) terms, and etc.. For convenience, when discussing one of those uncertainties, the other input parameters will be set as their central values.

The uncertainty caused different choice of charm quark mass is estimated by taking $m_c=1.5\pm0.1\,{\rm GeV}$. And we obtain
\begin{eqnarray}
\Delta \sigma|_{m_c}^{\rm Conv.}&=& ^{+0.073}_{-0.060}~\rm pb, \\
\Delta \sigma|_{m_c}^{\rm PMC}&=& ^{+0.091}_{-0.075}~\rm pb.
\end{eqnarray}

The uncertainty caused by $\Delta\alpha_s(M_Z)$ is obtained by using $\alpha_s(M_Z)=0.1179\pm0.0010$~\cite{ParticleDataGroup:2022pth}. And we obtain
\begin{eqnarray}
\Delta \sigma|_{\Delta\alpha_s(M_Z)}^{\rm Conv.}&=& ^{+0.013}_{-0.012}~\rm pb, \label{deltaalpha1}\\
\Delta \sigma|_{\Delta\alpha_s(M_Z)}^{\rm PMC}&=& ^{+0.022}_{-0.020}~\rm pb. \label{deltaalpha2}
\end{eqnarray}
Eqs.~(\ref{deltaalpha1}, \ref{deltaalpha2}) show that the PMC prediction is more sensitive to the value of $\Delta\alpha_s(M_Z)$. This is reasonable since the PMC uses the RGE-involved terms of the process to set the effective value of $\alpha_s$, thus its resultant pQCD series is more sensitive to the $\alpha_s$ running behavior that is governed by the RGE and the preciseness of the reference value $\alpha_s(M_Z)$.

Since the NLO-terms are sizable, it is helpful to have an estimation of the contributions from the UHO terms. For the purpose, we adopt the Pad\'{e} approximation approach (PAA)~\cite{Basdevant:1972fe, Samuel:1992qg, Samuel:1995jc}, which provides a practical way for promoting a finite series to an analytic function, to do such an estimation. Detailed procedures for such an estimation can be found in Ref.\cite{Du:2018dma}. For the present case, the estimated $\alpha_s^4$-terms (e.g. the NNLO-terms) of the conventional series and PMC series are
\begin{eqnarray}
\Delta \sigma|_{\rm High~order}^{\rm Conv.}&=&\pm c_2^2/c_1 a_s^4(2m_c) = \pm 0.080~\rm pb, \\
\Delta \sigma|_{\rm High~order}^{\rm PMC}&=&\pm r_{2,0}^2/r_{1,0}a_s^4(Q_*) = \pm 0.047~\rm pb. \label{pmcUN}
\end{eqnarray}

If including all those error sources, our final predictions for the NLO total cross section $e^+e^-\to J/\psi+c+\bar{c}$ are
\begin{eqnarray}
\sigma|_{\rm NLO}^{\rm Conv.}&=& 0.269^{+0.166}_{-0.124}~\rm pb, \\
\sigma|_{\rm NLO}^{\rm PMC}&=& 0.362^{+0.105}_{-0.091}~\rm pb,
\end{eqnarray}
where the net uncertainties are squared averages of those from the choice of renormalization scale $\mu_R\in[m_c, 4m_c]$, the error of charm quark mass $\Delta m_c$, the magnitude of $\Delta\alpha_s(M_Z)$, and the estimated NNLO contribution by the PAA. It shows that due to the improvement of pQCD series, the net uncertainty from all those error sources changes down from $108\%$ of the conventional series to $54\%$ of the PMC series.

To compare with the experimental data, we should also include the QED contribution as well as the two-photon contribution via the process $e^+e^-\to 2\gamma^* \to J/\psi+c+\bar{c}$. Since the experimental measurements are for the prompt $J/\psi+c+\bar{c}$ production, we should also consider the feed-down contributions from the higher excited charmonium states such as $e^+e^-\to \psi(2S)+c+\bar{c} \to J/\psi+c+\bar{c}+X$ and $e^+e^-\to \chi_{cJ}+c+\bar{c} \to J/\psi+c+\bar{c}+X$. By using $\Gamma(\psi(2S) \to e^+e^-)=5.53~{\rm KeV}$ and the branching ratio for the $\psi(2S) \to J/\psi X$ transition fraction $Br[\psi(2S) \to J/\psi X]=61.4\%$~\cite{ParticleDataGroup:2022pth}, we obtain the contribution from $e^+e^-\to \psi(2S)+c+\bar{c} \to J/\psi+c+\bar{c}+X$ is $0.371~\sigma_{\rm NLO}$. The other contributions have been discussed in Ref.~\cite{Zhang:2006ay}, which lead to a shift of $0.069$ pb to the total cross section. We will give the comparison with experimental data in the following section.

\subsection{Differential cross section}

The differential distributions contain more information than the integrated cross sections. Experimentally, the $J/\psi$ three-momentum and angular distributions for $e^+e^- \to J/\psi+c+\bar{c}$ have been measured by the Belle collaboration~\cite{Belle:2009bxr}. It is interesting to see the differential distributions of the process $e^+e^- \to J/\psi+c+\bar{c}$.

\begin{figure}[htb]
\center
\includegraphics[width=0.45\textwidth]{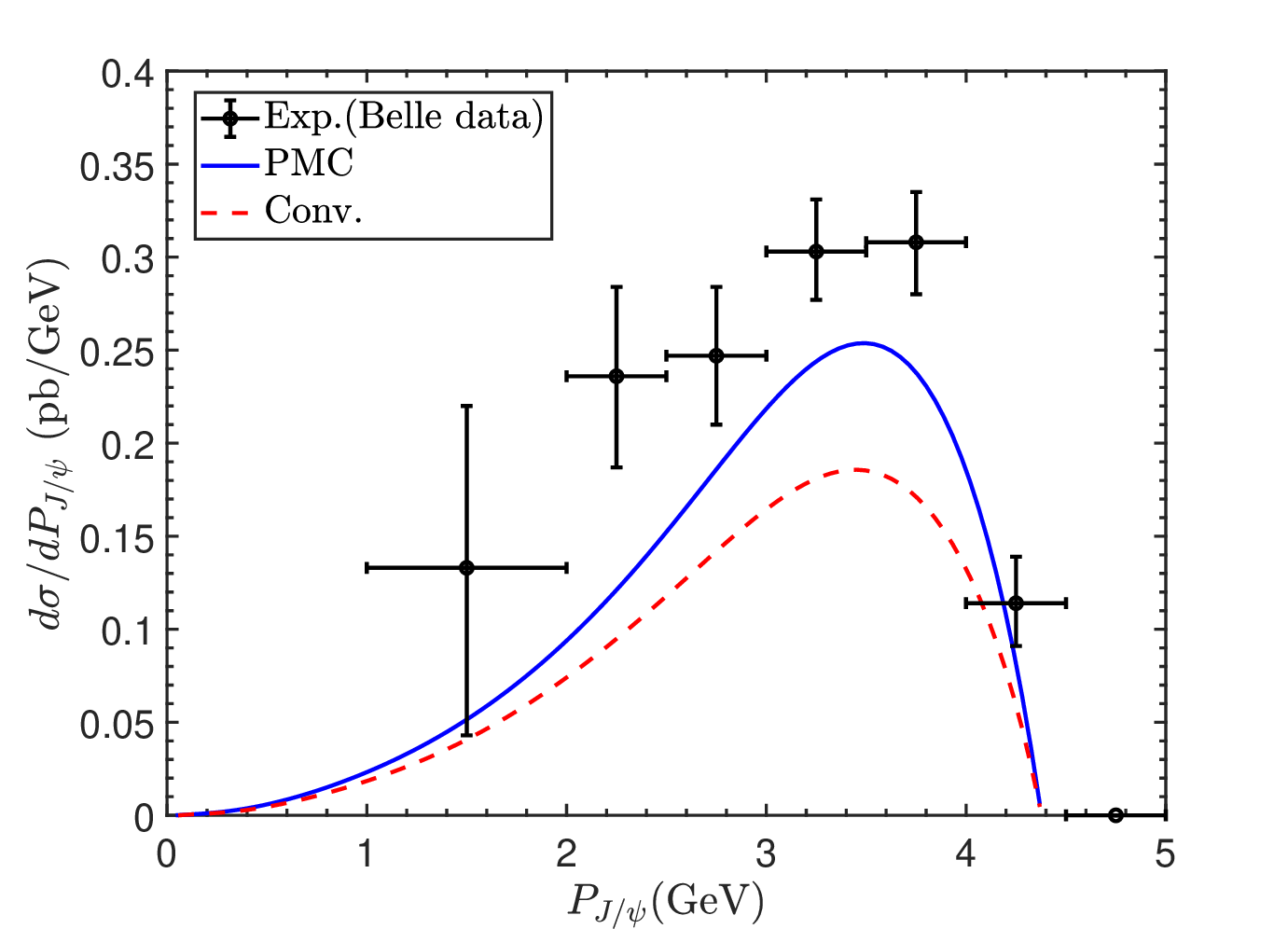}
\includegraphics[width=0.45\textwidth]{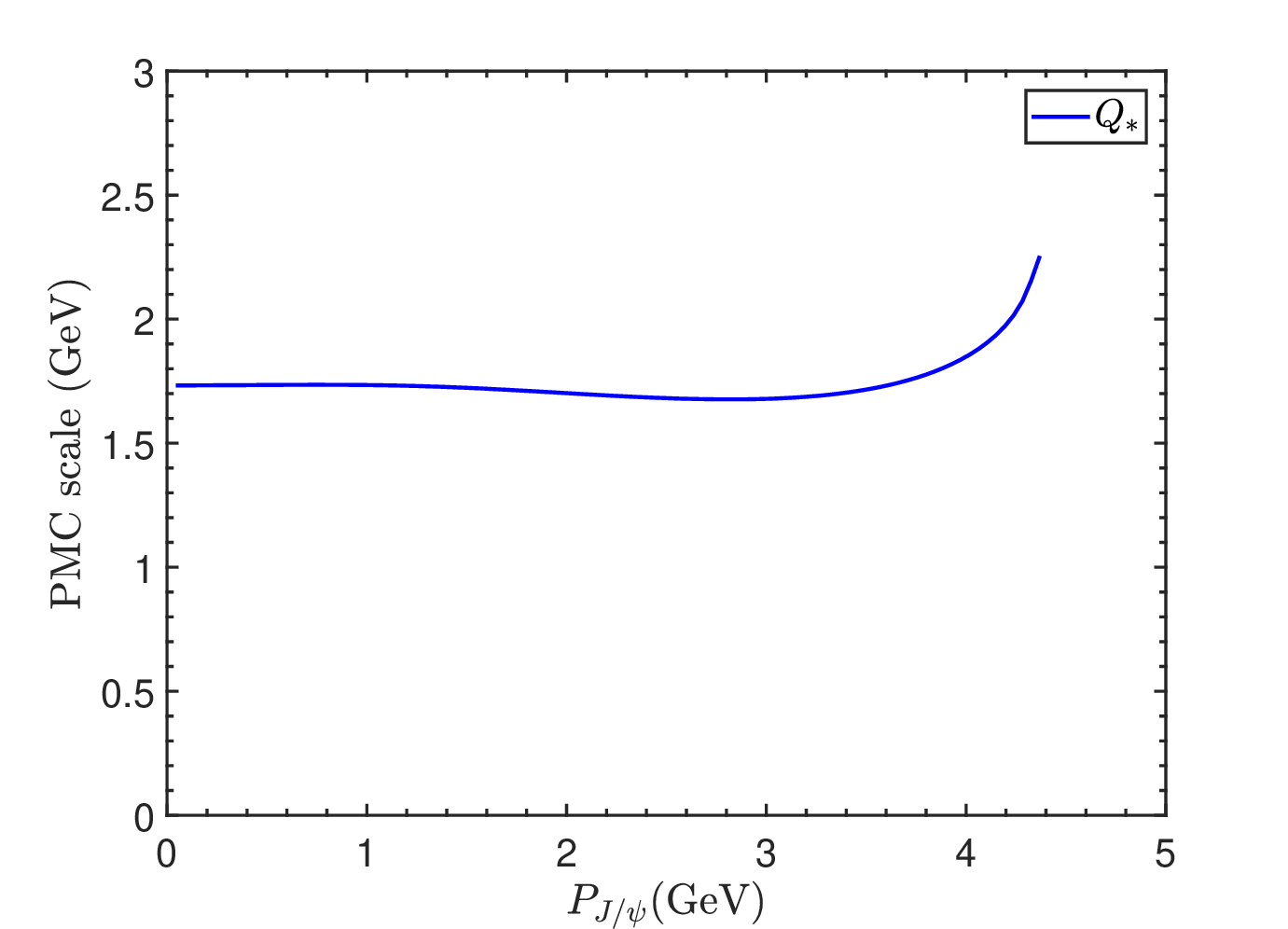}
\caption{The Left diagram shows the NLO differential cross sections $d\sigma/dP_{J/\psi}$ for $e^+e^- \to J/\psi+c+\bar{c}$ under conventional and PMC scale-setting approaches, respectively. Contributions from the feed-down of $\psi(2S)$ has been added to those two curves by multiplying a factor of 1.371. All the input parameters are set to be their central values and $\mu_R=2m_c$. The PMC scale $Q_*$ at each $P_{J/\psi}$ is shown in the Right diagram. }
\label{dsdp1}
\end{figure}

We present the NLO differential cross sections $d\sigma/dP_{J/\psi}$ for $e^+e^- \to J/\psi+c+\bar{c}$ under the conventional and PMC scale-setting approaches in the left diagram of Fig.~\ref{dsdp1}, where the experimental data has been also shown for comparison. It shows that the shape of the $J/\psi$ three-momentum distribution under the PMC scale-setting approach agrees with the experimental data better than the conventional one. At each kinematic point, the effective momentum flow, e.g. the PMC scale $Q_*$, may be different. We present the PMC effective scale $Q_*$ versus $P_{J/\psi}$ in the right diagram of Fig.~\ref{dsdp1}. In most $P_{J/\psi}$-range, the $Q_*$ has close values, but it becomes larger at high $P_{J/\psi}$, so we cannot set a unified $Q_*$ for the whole phase space for achieving precise prediction.

\begin{figure}[htb]
\center
\includegraphics[width=0.45\textwidth]{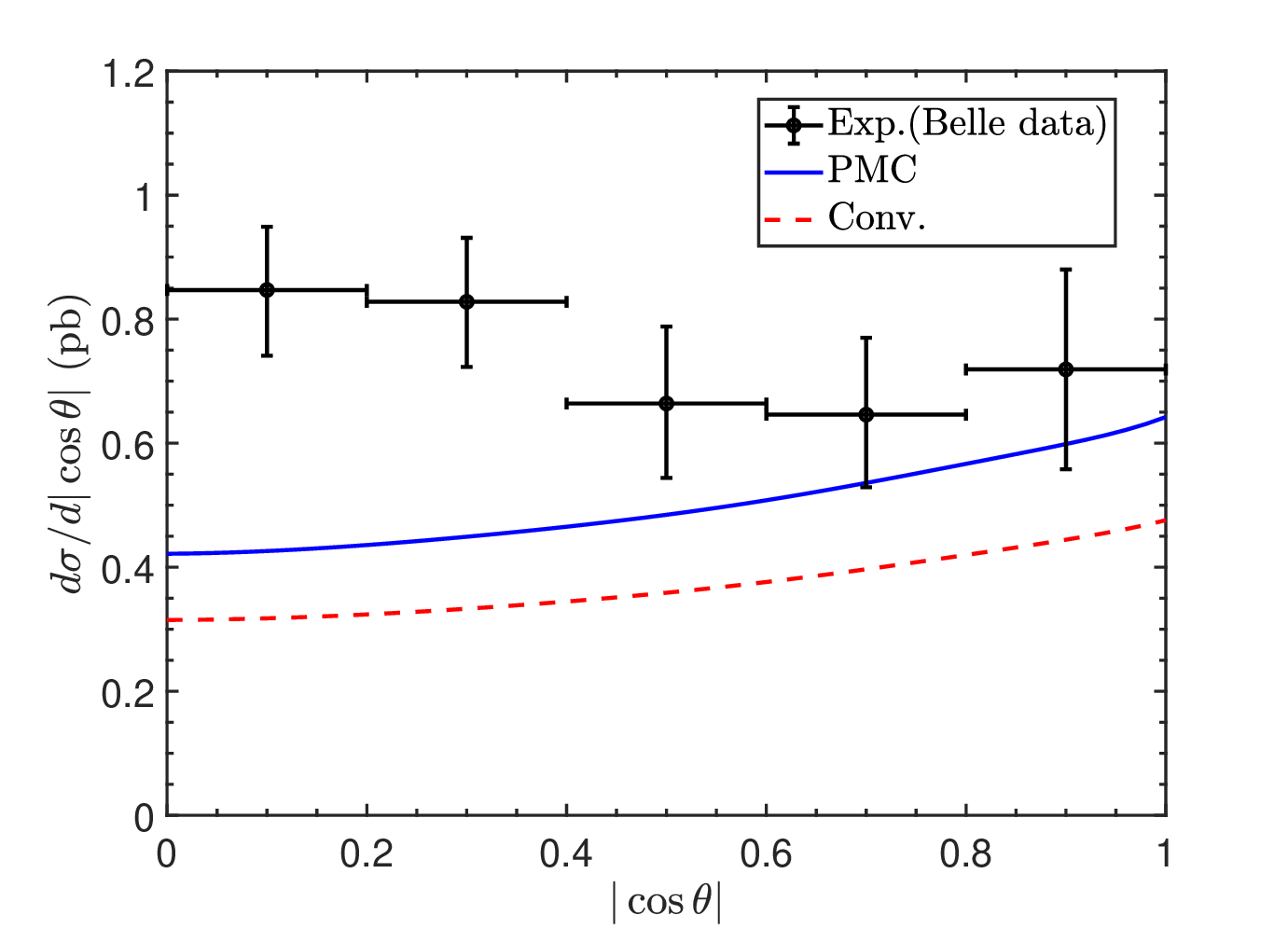}
\includegraphics[width=0.45\textwidth]{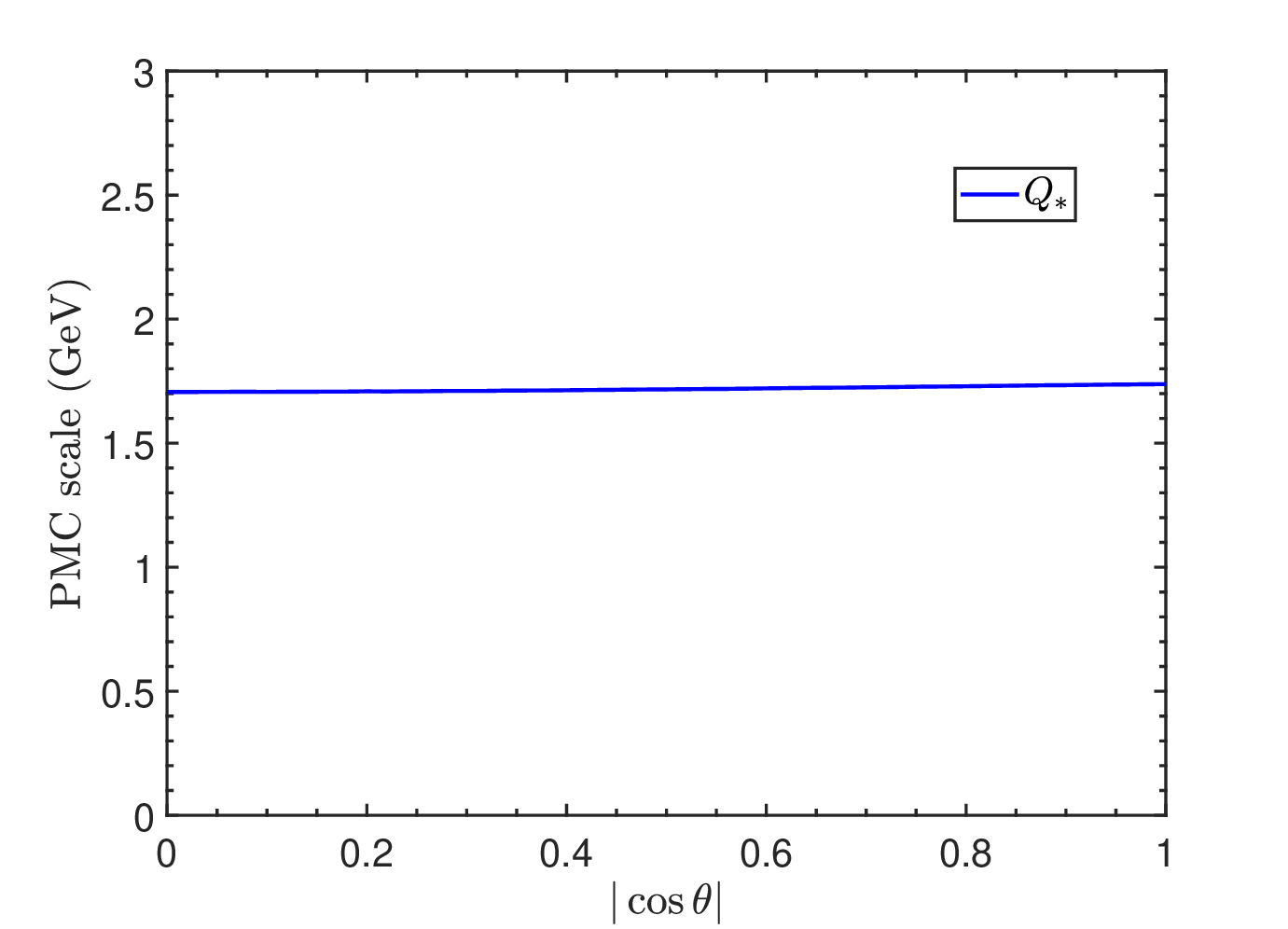}
\caption{The NLO differential cross sections $d\sigma/d|\cos \theta|$ for $e^+e^- \to J/\psi+c+\bar{c}$ under conventional and PMC scale-setting approaches, respectively. Contributions from the feed-down of $\psi(2S)$ has been added to those two curves by multiplying a factor of 1.371. All the input parameters are set to be their central values and $\mu_R=2m_c$. The PMC scale $Q_*$ at each $|\cos \theta|$ is shown in the Right diagram. }
\label{dsdcos}
\end{figure}

\begin{figure}[htb]
\center
\includegraphics[width=0.45\textwidth]{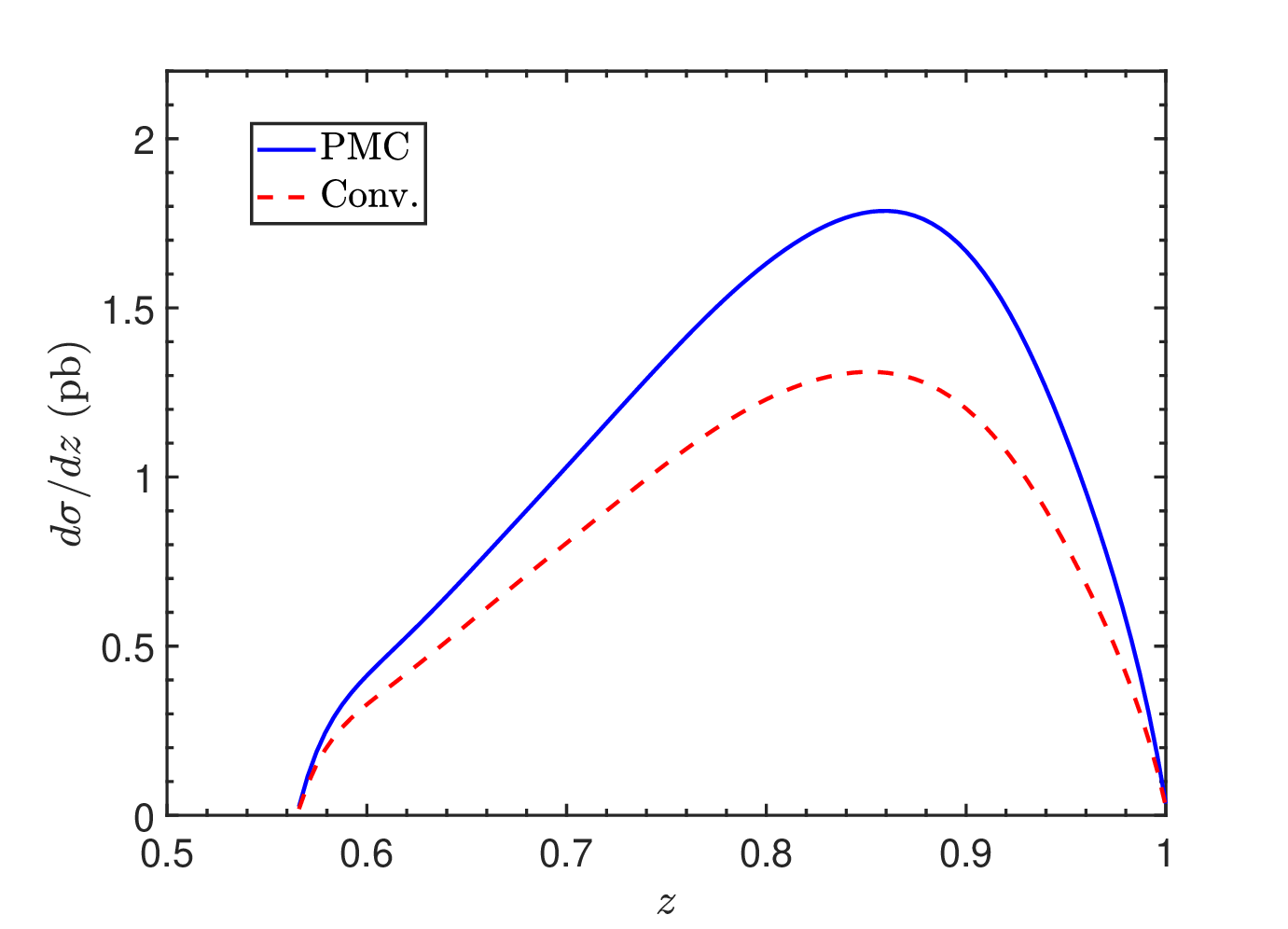}
\includegraphics[width=0.45\textwidth]{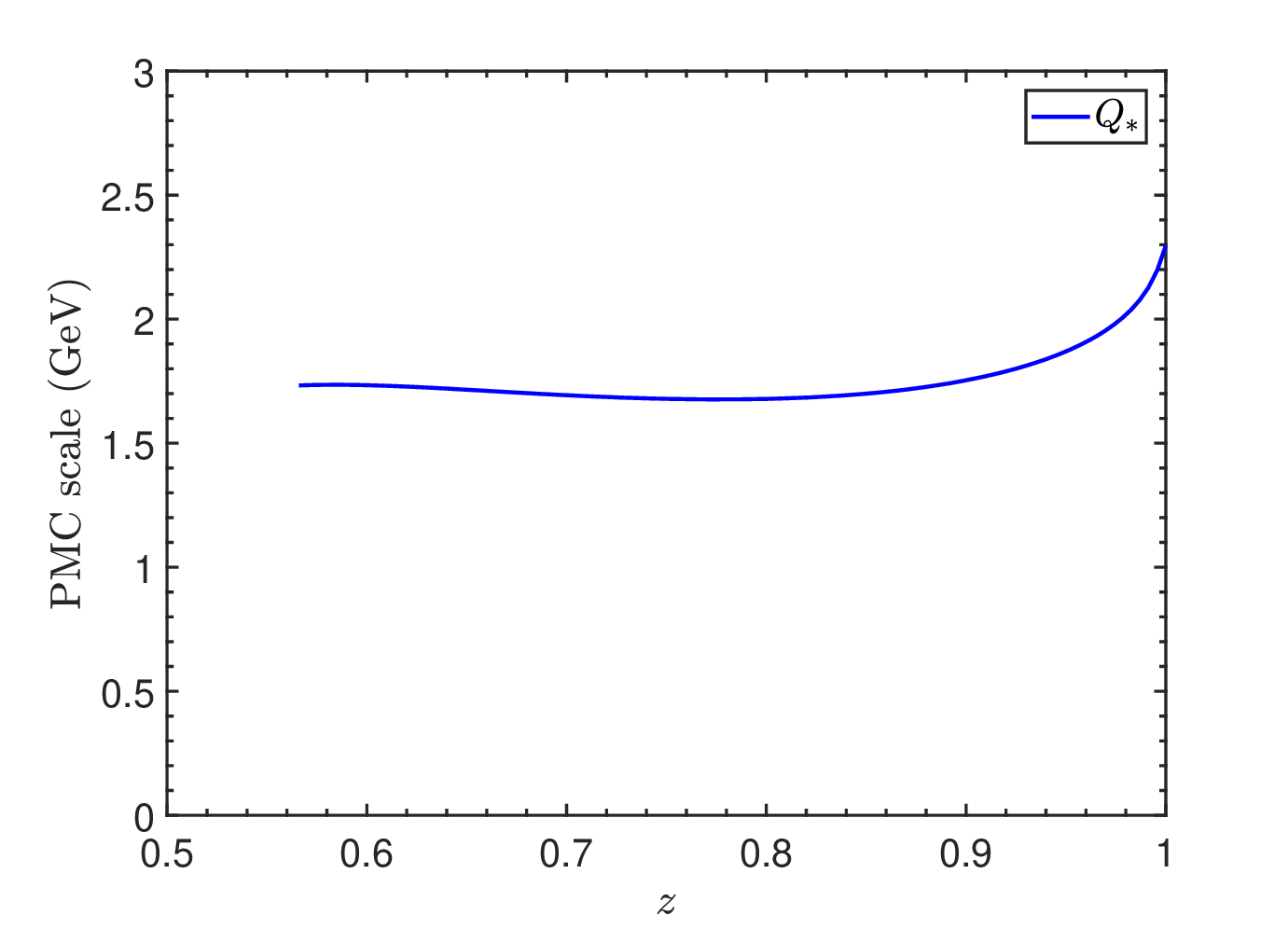}
\caption{The NLO differential cross section $d\sigma/dz$ for $e^+e^- \to J/\psi+c+\bar{c}$ under conventional and PMC scale-setting approaches, respectively. Contributions from the feed-down of $\psi(2S)$ has been added to those two curves by multiplying a factor of 1.371. All the input parameters are set to be their central values and $\mu_R=2m_c$. The PMC scale $Q_*$ at each $z$ is shown in the Right diagram. }
\label{dsdz}
\end{figure}

Similarly, we present the NLO differential cross sections $d\sigma/d|\cos \theta|$ for $e^+e^- \to J/\psi+c+\bar{c}$ under conventional and PMC scale-setting approaches in Fig.~\ref{dsdcos}, respectively. The use of PMC raises the production rate, but will not change the arising trends of $d\sigma/d|\cos \theta|$ versus $|\cos \theta|$, i.e. it is estimated that more events should be accumulated when the produced $J/\psi$ moves forward or backward along with the $e^+ e^-$ collision direction. But the data indicate that there may have more $J/\psi$ moves perpendicularly to the $e^+ e^-$ collision direction. So more precise data or the NNLO and higher-order corrections may have some help for solving this discrepancy. We also present the NLO differential cross sections $d\sigma/dz$ for the process $e^+e^- \to J/\psi+c+\bar{c}$ in Fig.~\ref{dsdz}. Here $z$ is the energy fraction carrying by the produced $J/\psi$, which is defined as $z \equiv 2\,k\cdot p_{1}/s$, where $k$ is the sum of the momenta of the initial electron and positron. Fig.~\ref{dsdz} shows that the magnitude of $d\sigma/dz$ is increased at small $z$ values and decreased at higher $z$ values.

\begin{figure}[htbp]
\center
\includegraphics[width=0.45\textwidth]{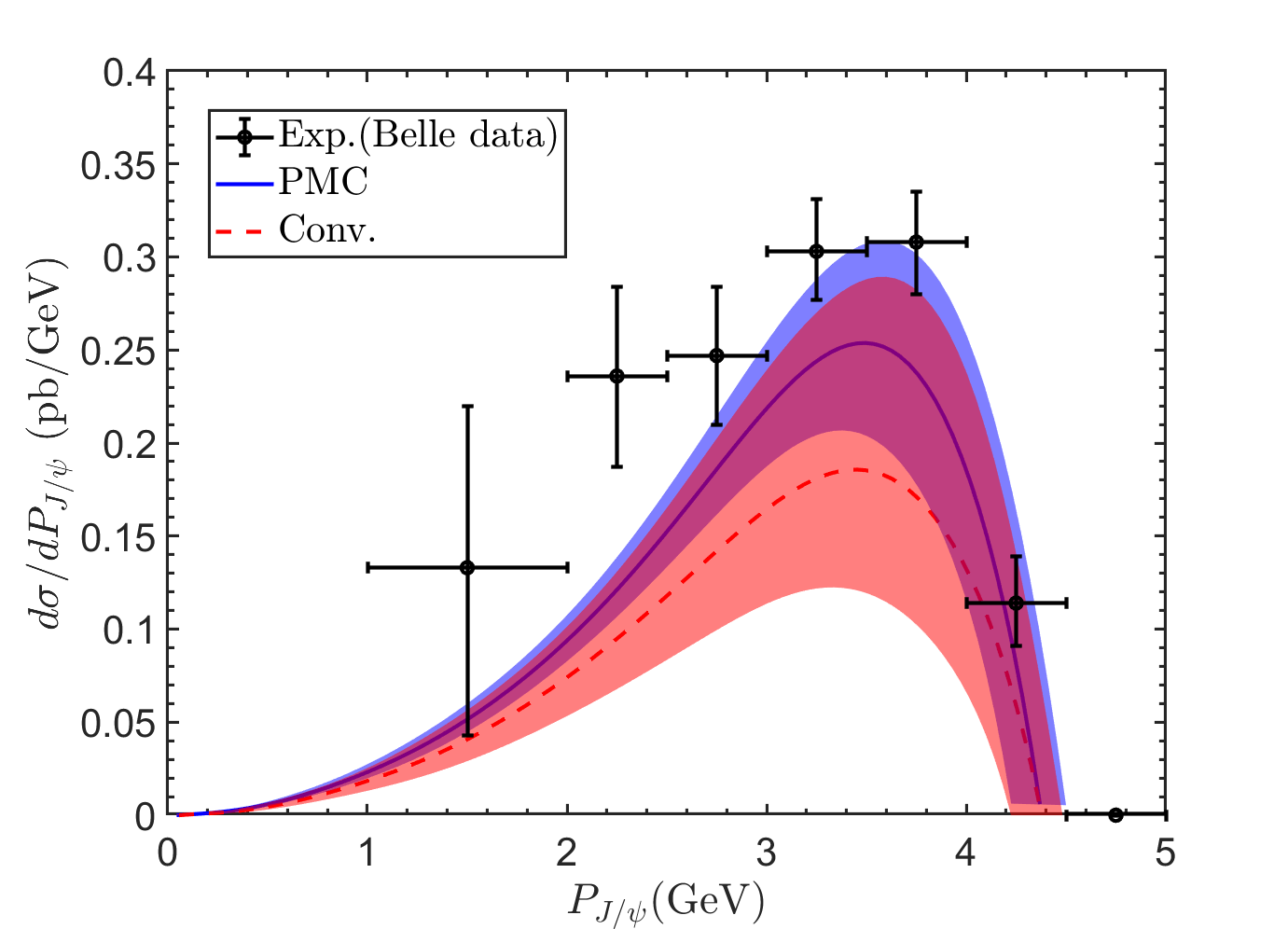}
\caption{The NLO differential cross sections $d\sigma/dP_{J/\psi}$ for $e^+e^- \to J/\psi+c+\bar{c}$ under the conventional and PMC scale-setting approaches. The contribution from the feed-down of $\psi(2S)$ has been added to all curves by multiplying a factor of 1.371. The error bands are squared averages of the errors caused by taking $\mu_R \in [m_c, 4m_c]$ and $m_c=1.5\pm0.1 {\rm GeV}$. }
\label{dsdp11}
\end{figure}

\begin{figure}[htbp]
\center
\includegraphics[width=0.45\textwidth]{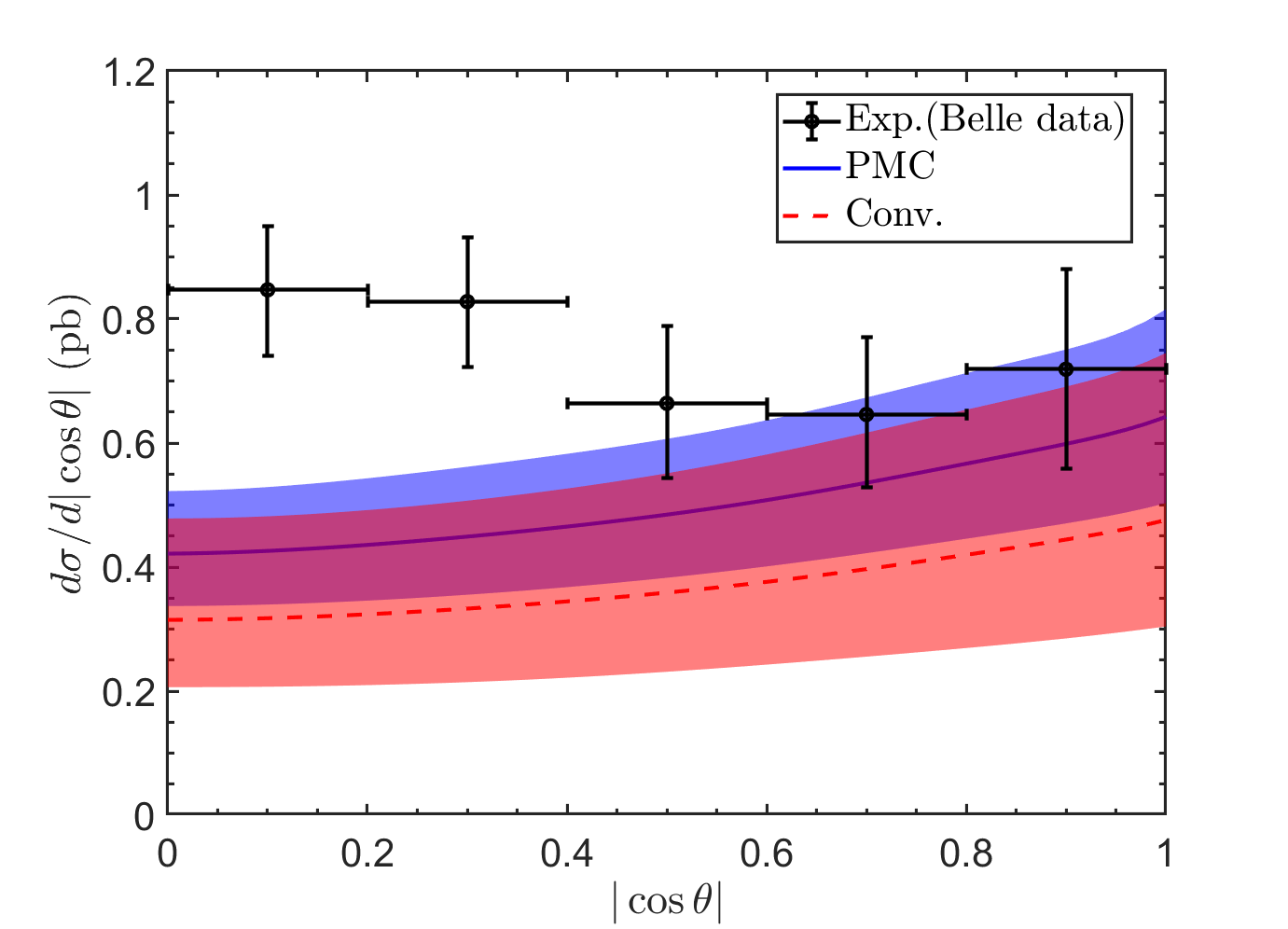}
\caption{The NLO differential cross sections $d\sigma/d|\cos \theta|$ for $e^+e^- \to J/\psi+c+\bar{c}$ under the conventional and PMC scale-setting approaches. The contribution from the feed-down of $\psi(2S)$ has been added to all curves by multiplying a factor of 1.371. The error bands are squared averages of the errors caused by taking $\mu_R \in [m_c, 4m_c]$ and $m_c=1.5\pm0.1 {\rm GeV}$. }
\label{dsdcos1}
\end{figure}

\begin{figure}[htbp]
\center
\includegraphics[width=0.45\textwidth]{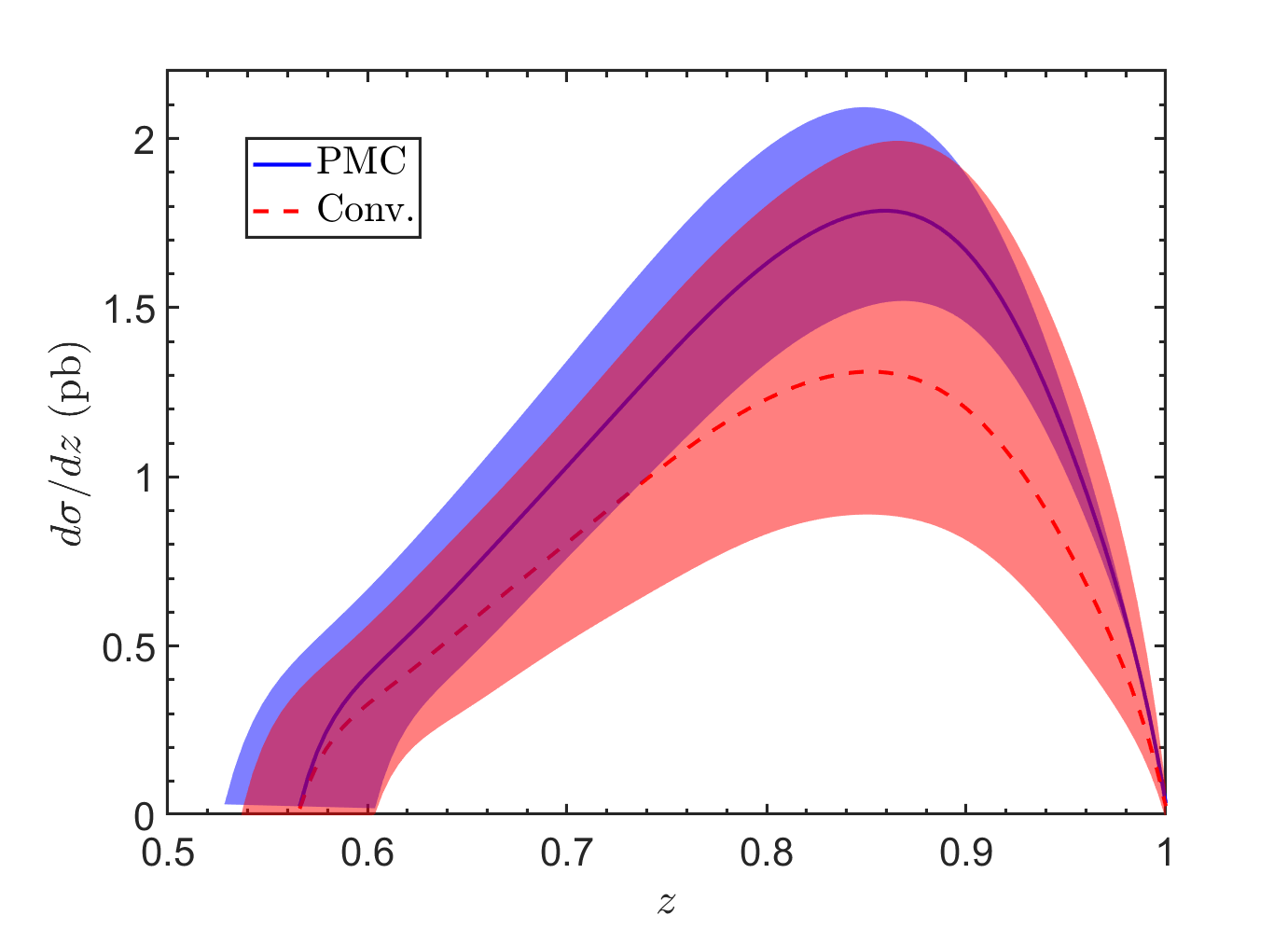}
\caption{The NLO differential cross sections $d\sigma/dz$ for $e^+e^- \to J/\psi+c+\bar{c}$ under the conventional and PMC scale-setting approaches. The contribution from the feed-down of $\psi(2S)$ has been added to all curves by multiplying a factor of 1.371. The error bands are squared averages of the errors caused by taking $\mu_R \in [m_c, 4m_c]$ and $m_c=1.5\pm0.1 {\rm GeV}$. }
\label{dsdz1}
\end{figure}

As a final remark, we show the uncertainties of the NLO differential cross sections $d\sigma/dP_{J/\psi}$, $d\sigma/d|\cos \theta|$, and $d\sigma/dz$ under conventional and PMC scale-setting approaches in Figs.~\ref{dsdp11}, \ref{dsdcos1}, and \ref{dsdz1}, respectively. Here for definiteness, we have set $\alpha_s(M_Z)=0.1179$. The error bands of those diagrams are squared averages of the errors caused by taking $\mu_R \in [m_c, 4m_c]$ and $m_c=1.5\pm0.1 {\rm GeV}$.

\subsection{Self-consistency of PMC to integrated and differential cross sections}

In this subsection, we give a first self-consistency discussion on the application of PMC scale-setting approach to either the total cross-section or the differential cross-section. Usually, the PMC is applied for dealing with total cross section, fixing the correct magnitude of $\alpha_s$ via the RGE and then determining the overall effective momentum flow of the process. For the purpose, we set all the input parameters to be their central values. Using the series (\ref{beta}) for total cross section, we have obtained $Q_*=1.72~{\rm GeV}$. As has been shown in the Right diagrams of Figs. \ref{dsdp1}, \ref{dsdcos} and \ref{dsdz}, the PMC can also be applied for dealing with differential cross sections via the same way, which then sets the correct magnitude of $\alpha_s$ at each point of $P_{J/\psi}$, $|\cos\theta|$, $z$, and etc.. The PMC scale at the level of differential cross section represents the effective momentum flow at each point, which is derived by the resummation of the RGE-involved $\{\beta_i\}$-terms at this specific point. Thus it is generally different for different choice of variable and for different value of this particular variable. It is interesting to know whether the weighted average of the PMC scales at each point will lead to the same overall PMC scale. By using 100 points as a basis and using the differential value at each point as its weight, we numerically obtain $Q_*|^{P_{J/\psi}}=1.73~{\rm GeV}$, $Q_*|^{|\cos\theta|}=1.72~{\rm GeV}$, and $Q_*|^z=1.73~{\rm GeV}$, which are derived by using the $P_{J/\psi}$-distribution, $|\cos\theta|$-distribution and $z$-distribution, respectively.

After integrating over the differential variable within its allowable range, one will naturally obtain its total cross section. This is surely case for conventional pQCD series, in which the renormalization scale is taken to be the same for either total cross section or differential cross section. For the present case, we have $\sigma_{\rm NLO}|^{P_{J/\psi}}=0.269$~pb, $\sigma_{\rm NLO}|^{|\cos\theta|}=0.270$~pb, and $\sigma_{\rm NLO}|^z=0.269$~pb, which are derived by using the $P_{J/\psi}$-distribution, $|\cos\theta|$-distribution and $z$-distribution, respectively. All of them are in good agreement with the above derived total cross section $\sigma_{\rm NLO}|_{\rm Conv.}=0.269$~pb.

\begin{table}[htb]
\center
\begin{tabular}{c c c c c}
\hline
  & ~~$\sigma_{\rm NLO}|^{P_{J/\psi}}$~~ & ~~$\sigma_{\rm NLO}|^{|\cos\theta|}$~~ & ~~$\sigma_{\rm NLO}|^z$~~ & ~~$\sigma_{\rm NLO}$~~  \\
\hline
${\rm PMC}$   & $0.359$ & $0.365$ & $0.359$ & $0.362$   \\
\hline
\end{tabular}
\caption{The PMC predictions for total cross section of $e^+e^-\to J/\psi+c+\bar{c}$ (in unit: pb) by starting from either differential cross section or total cross section up to NLO QCD corrections, respectively. The upper indexes ${P_{J/\psi}}$, ${|\cos\theta|}$ and ${z}$ denote the ones derived from integrating over the differential cross sections $d\sigma/dP_{J/\psi}$, $d\sigma/d|\cos\theta|$ and $d\sigma/dz$, respectively. $\sigma_{\rm NLO}$ is calculated from the total cross section series (\ref{pmc}). }
\label{tb.diff}
\end{table}

We show the PMC NLO predictions of the total cross sections in Table \ref{tb.diff}, which are derived either from the differential series ($d\sigma/dP_{J/\psi}$, $d\sigma/d|\cos\theta|$ and $d\sigma/dz$) or from the series (\ref{pmc}). The upper indexes ${P_{J/\psi}}$, ${|\cos\theta|}$ and ${z}$ denote the ones derived from integrating over the differential cross sections $d\sigma/dP_{J/\psi}$, $d\sigma/d|\cos\theta|$ and $d\sigma/dz$, respectively. Table \ref{tb.diff} shows that all those PMC predictions are consistent with each other. This shows that the PMC predictions can also get the self-consistent results for the differential and total cross sections. But there are differences in the details of getting the same conclusion. There are slight differences, which are less than $\pm 1\%$, among the results obtained by integrating over the differential distributions with the result obtained from the total series (\ref{pmc}). Such slight difference is reasonable, 1) the differences are due to the resummation of different types of $\{\beta_i\}$-terms, and are caused by the unknown $\{\beta_i\}$-types from the unknown NNLO and higher order terms; 2) The effects of the differences of the $\{\beta_i\}$-terms will result in the differences of the PMC scales, which generally suffer from both exponential and $\alpha_s$ suppressions. So the net value for such difference could be small. For the present case of $e^+e^- \to J/\psi+c+\bar{c}$, we observe that the differences are about $\pm 0.003$~pb, which are higher-order effects and are under well control~\footnote{For some cases, certain large logarithmic terms may appear in boundary region of the phase space, which may spoil the convergent behavior of the perturbative series for some specific phase-space points. And special resummation, such as the threshold resummation~\cite{Yan:2023mjj}, need to be done so as to achieve a more reliable PMC prediction.}. And for the present case, their magnitudes are much smaller than the predicted NNLO magnitude $\pm 0.047$~pb. Thus we can conclude that the PMC can be applied to any pQCD calculable observable via a self-consistent way of perturbative theory, and it can achieve more precise fixed-order perturbative predictions that are free of renormalization scheme-and-scale ambiguities.

\section{Summary}
\label{secSum}

In the present paper, we have studied the total and differential cross sections of $e^+e^- \to J/\psi+c+\bar{c}$ up to NLO QCD corrections. After including the NLO QCD corrections, the perturbative series under conventional scale-setting approach is still highly scale dependent, e.g. the net scale error is still $\sim 74\%$ for $\mu_R\in [m_c, 4 m_c]$. While after applying the PMC, we achieve a scale-invariant and more convergent prediction, and the NLO terms will enhance the total cross section by $\sim 43\%$. The higher excited states such as $\psi(2S)$ and $\chi_{cJ}$ may decay into $J/\psi$ with high probability, and by taking those contributions into consideration, we obtain the prompt total cross sections for $e^+e^- \to J/\psi+c+\bar{c}$, i.e.,
\begin{eqnarray}
\sigma|_{\rm prompt}^{\rm Conv.}&=& 0.438^{+0.228}_{-0.170}~\rm pb, \\
\sigma|_{\rm prompt}^{\rm PMC}&=& 0.565^{+0.144}_{-0.125}~\rm pb,
\end{eqnarray}
which are obtained by using $\sigma_{\rm prompt}=1.371\sigma_{\rm NLO}+0.069$~pb. Here the uncertainties are squared averages of those from the choice of renormalization scale $\mu_R\in[m_c, 4m_c]$, the error of charm quark mass $\Delta m_c$, the magnitude of $\Delta\alpha_s(M_Z)$, and the estimated NNLO contribution by the PAA.

\begin{figure}[htb]
\center
\includegraphics[width=0.45\textwidth]{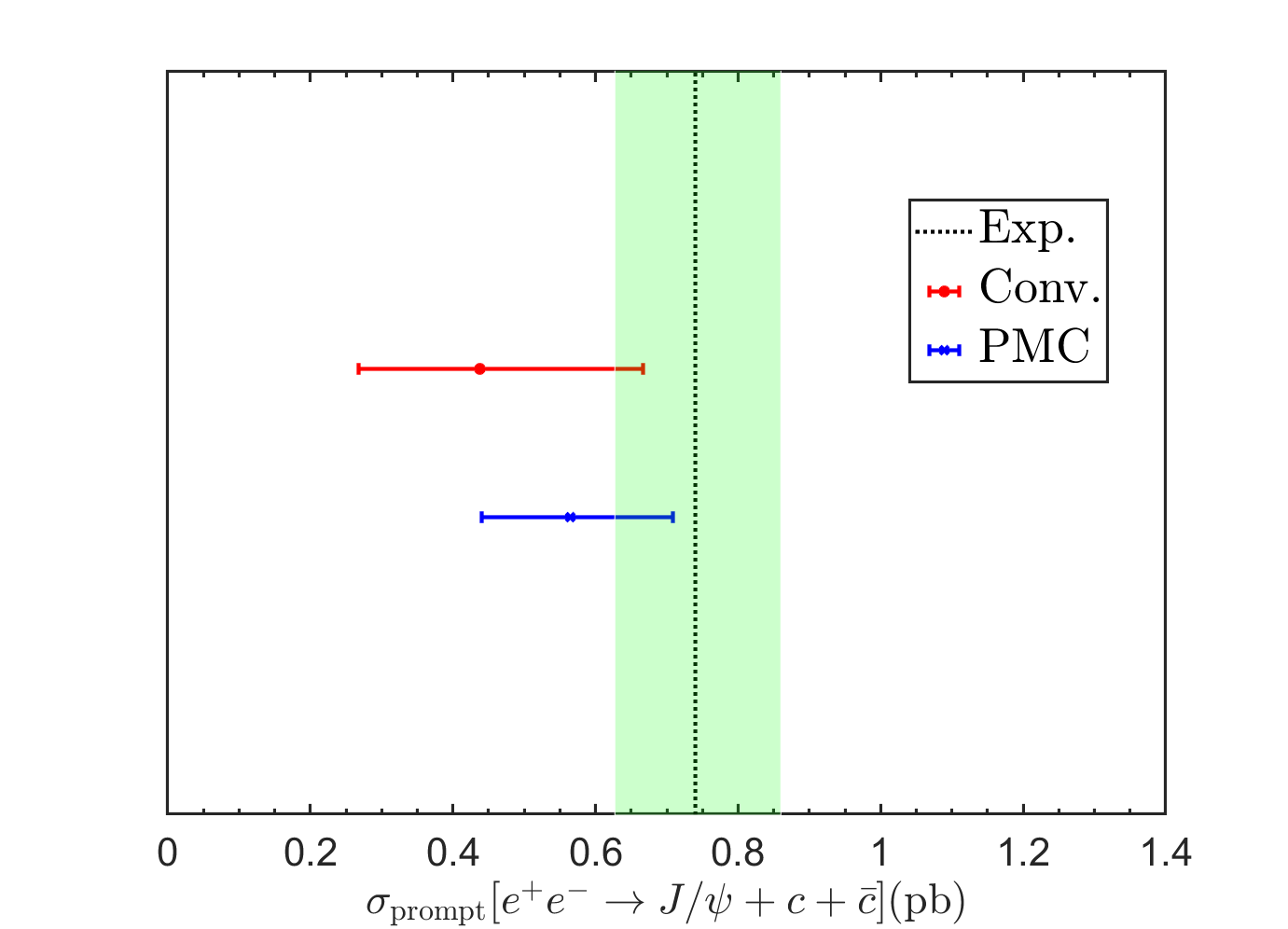}
\caption{The prompt total cross sections of $e^+e^- \to J/\psi+c+\bar{c}$ under conventional and PMC scale-setting approaches, respectively. The Belle measured value~\cite{Belle:2009bxr} is given as a comparison. } \label{error}
\end{figure}

In Fig.~\ref{error}, we present a comparison of the prompt total cross section of $e^+e^- \to J/\psi+c+\bar{c}$ with the Belle measured value, $\sigma^{\rm Exp.}(e^+e^- \to J/\psi+c+\bar{c})=0.74^{+0.113}_{-0.120}$ pb, whose errors are squared averages of the systematic and statistical errors given in Eq.(\ref{belleII}). Fig.~\ref{error} shows that a better agreement with the experimental data can be achieved by applying the PMC.

After applying the PMC scale-setting approach, one can determine the correct magnitude of the effective $\alpha_s$, generally resulting in a more convergent pQCD series that is free of divergent renormalon terms. Thus, a reliable, self-consistent and precise pQCD prediction can be achieved. We have also done a first consistency test of the PMC approach to deal with the total and differential observables. As a byproduct, the typical momentum flow of the process may be small in specific phase space, the consistency property of the PMC production makes its differential series be a useful platform for testing the correctness of the low-energy $\alpha_s$ models.

\hspace{2cm}

\noindent {\bf Acknowledgments:} We thank Stanley J. Brodsky for helpful discussions. This work was supported in part by the National Natural Science Foundation of China under Grant Nos. 12175025, 12347101, 12247129, 11975242, and 12135013; by the Graduate Research and Innovation Foundation of Chongqing, China under Grant No. ydstd1912; by the Foundation of Chongqing Normal University under Grant No. 24XLB015; by the Fundamental Research Funds for the Central Universities under Grant No. 2024CDJXY022.

\hspace{2cm}

\end{document}